\documentclass[fleqn,usenatbib,letterpaper]{mn2e}
\usepackage{graphicx}
\usepackage{amsmath}
\def\jnl@style{\it}
\def\aaref@jnl#1{{\jnl@style#1}}

\def\aaref@jnl#1{{\jnl@style#1}}

\def\aj{\aaref@jnl{AJ}}                   
\def\araa{\aaref@jnl{ARA\&A}}             
\def\apj{\aaref@jnl{ApJ}}                 
\def\apjl{\aaref@jnl{ApJ}}                
\def\apjs{\aaref@jnl{ApJS}}               
\def\ao{\aaref@jnl{Appl.~Opt.}}           
\def\apss{\aaref@jnl{Ap\&SS}}             
\def\aap{\aaref@jnl{A\&A}}                
\def\aapr{\aaref@jnl{A\&A~Rev.}}          
\def\aaps{\aaref@jnl{A\&AS}}              
\def\azh{\aaref@jnl{AZh}}                 
\def\baas{\aaref@jnl{BAAS}}               
\def\jrasc{\aaref@jnl{JRASC}}             
\def\memras{\aaref@jnl{MmRAS}}            
\def\mnras{\aaref@jnl{MNRAS}}             
\def\pra{\aaref@jnl{Phys.~Rev.~A}}        
\def\prb{\aaref@jnl{Phys.~Rev.~B}}        
\def\prc{\aaref@jnl{Phys.~Rev.~C}}        
\def\prd{\aaref@jnl{Phys.~Rev.~D}}        
\def\pre{\aaref@jnl{Phys.~Rev.~E}}        
\def\prl{\aaref@jnl{Phys.~Rev.~Lett.}}    
\def\pasp{\aaref@jnl{PASP}}               
\def\pasj{\aaref@jnl{PASJ}}               
\def\qjras{\aaref@jnl{QJRAS}}             
\def\skytel{\aaref@jnl{S\&T}}             
\def\solphys{\aaref@jnl{Sol.~Phys.}}      
\def\sovast{\aaref@jnl{Soviet~Ast.}}      
\def\ssr{\aaref@jnl{Space~Sci.~Rev.}}     
\def\zap{\aaref@jnl{ZAp}}                 
\def\nat{\aaref@jnl{Nature}}              
\def\iaucirc{\aaref@jnl{IAU~Circ.}}       
\def\aplett{\aaref@jnl{Astrophys.~Lett.}} 
\def\apspr{\aaref@jnl{Astrophys.~Space~Phys.~Res.}}
\def\bain{\aaref@jnl{Bull.~Astron.~Inst.~Netherlands}} 
\def\fcp{\aaref@jnl{Fund.~Cosmic~Phys.}}  
\def\gca{\aaref@jnl{Geochim.~Cosmochim.~Acta}}   
\def\grl{\aaref@jnl{Geophys.~Res.~Lett.}} 
\def\jcp{\aaref@jnl{J.~Chem.~Phys.}}      
\def\jgr{\aaref@jnl{J.~Geophys.~Res.}}    
\def\jqsrt{\aaref@jnl{J.~Quant.~Spec.~Radiat.~Transf.}}
\def\memsai{\aaref@jnl{Mem.~Soc.~Astron.~Italiana}}
\def\nphysa{\aaref@jnl{Nucl.~Phys.~A}}   
\def\physrep{\aaref@jnl{Phys.~Rep.}}   
\def\physscr{\aaref@jnl{Phys.~Scr}}   
\def\planss{\aaref@jnl{Planet.~Space~Sci.}}   
\def\procspie{\aaref@jnl{Proc.~SPIE}}   

\begin{document}
\title[Analytical Solutions to the Mass-Anisotropy Degeneracy]{Analytical Solutions to the Mass-Anisotropy Degeneracy with Higher Order Jeans Analysis: A General Method}
\author[T. Richardson \& M. Fairbairn]{Thomas Richardson\thanks{thomas.d.richardson@kcl.ac.uk}$^{1}$, Malcolm Fairbairn\thanks{malcolm.fairbairn@kcl.ac.uk}$^{1}$\\$^{1}$Physics, Kings College London, Strand, London WC2R 2LS, UK}
\maketitle
\begin{abstract}
The Jeans analysis is often used to infer the total density of a system by relating the velocity moments of an observable tracer population to the underlying gravitational potential. This technique has recently been applied in the search for Dark Matter in objects such as dwarf spheroidal galaxies where the presence of Dark Matter is inferred via stellar velocities. A precise account of the density is needed to constrain the expected gamma ray flux from DM self-annihilation and to distinguish between cold and warm dark matter models. Unfortunately the traditional method of fitting the second order Jeans equation to the tracer dispersion suffers from an unbreakable degeneracy of solutions due to the unknown velocity anisotropy of the projected system. To tackle this degeneracy one can appeal to higher moments of the Jeans equation.  By introducing an analog to the Binney anisotropy parameter at fourth order, $\beta'$ we create a framework that encompasses all solutions to the fourth order Jeans equations rather than the restricted range imposed by the separable augmented density. The condition $\beta' = f(\beta)$ ensures that the degeneracy is lifted and we interpret the separable augmented density system as the order-independent case $\beta'=\beta$. For a generic choice of $\beta'$ we present the line of sight projection of the fourth moment and how it could be incorporated into a joint likelihood analysis of the dispersion and kurtosis. The framework is then extended to all orders such that constraints may be placed to ensure a physically positive distribution function.  Having presented the mathematical framework, we then use it to make preliminary analyses of simulated dwarf spheroidal data leading to interesting results which strongly motivate further study.
\end{abstract}


\section{Introduction}
The favoured $\Lambda$CDM model of cosmology is consistent with a large invisible non-baryonic component of matter. To infer its existence astronomers thus look for the gravitational effect of its significant mass upon luminous tracer objects or for the observable products of DM annihilation and/or decay such as gamma rays \citep{gunn78,stecker78} which have been used in recent searches \citep[e.g][]{abdo} for dark matter. In both instances the density distribution of the system is critical with the Earth-incident flux of annihilation products dependent not only on model-dependent properties derived from particle physics \citep[see e.g.][]{pieri2009} but also on the square of the density distribution of dark matter within the source.  This is encoded in what is known as the \textit{astrophysical J-factor} which can be written \citep{walker2011},
\begin{equation}\label{jfac}
J(\theta_{\text{int}}) = \frac{4\pi}{d^2}\int^{\theta_{\text{int}}d}_{0}r^2\rho^{2}_{\text{dm}}(r)\text{d}r
\end{equation}
where $d$ is the distance to the source, $\rho_{\text{dm}}(r)$ is the local density of dark matter and $\theta_{\text{int}}$ is the integration angle which is related to a given solid angle of the source via $\Delta\Omega=2\pi(1-\cos\theta_{\text{int}})$.  The quadratic dependence upon the density introduces a very significant and DM model independent contribution to the flux which makes the choice of astrophysical source critically \citep{penarrubia} important for optimising DM searches. Though one might expect that the galactic center would provide the strongest signal, the strong and chaotic astrophysical backgrounds make it arguably less favourable than dwarf spheroidal galaxies (dSphs) of the local group which have been identified as having a large mass-to-light discrepancy \citep{aaronson} suitable for dark matter searches \citep{Lake, evans04}. This, in conjunction with their relative proximity to earth, makes them natural laboratories for DM and in recent years it has been possible to obtain samples of stellar positions and velocities \citep[e.g][]{walkdat} that are large enough for statistical treatment. Since the typical angular resolution of gamma ray telescopes is larger than the angular size on the sky of dwarf spheroidal galaxies \citep{abdo}, it turns out that the J-factor is not extremely sensitive to the distribution of dark matter in the core of dwarf spheroidals, although in the event of a signal being observed we would ideally want a better indication of the J-factors than the range of current estimates which vary over about an order of magnitude.

Another reason why it is important to study the centre of dwarf spheroidals is to probe another aspect of dark matter, namely its primordial velocity.  In the cold dark matter hypothesis the kinetic energy of dark matter at the start of structure formation is some very small fraction of its rest mass energy \citep{stefan, green} and the smallest structures above this free streaming scale are the first to form.  In models of hot dark matter \citep[see e.g.][]{zeldovich} dark matter begins completely relativistic and the largest structures form first.  A (rather finely tuned) compromise between these two extremes is the idea of warm dark matter \citep{wdm} where dark matter is not created with highly relativistic velocities, but nevertheless with significant velocities, meaning that the normal growth pattern of cold dark matter proceeds only above a length scale related to the free streaming length corresponding to the initial velocity.  This idea has been invoked to explain the lack of predicted satellites of the Milky Way \citep{wherearethey} as well as some interpretations of tracer populations in dwarf spheroidals wherein it is argued that dark matter halos possess a significant core, possibly due to some inherent initial kinetic energy \citep{gilmore}. At the same time, the importance of the role of baryons upon dark matter density in the core of halos is becoming increasingly clear \citep{governato}. Whatever the underlying physics, it is clear that we would like to be able to interpret the stellar velocity dispersion in such objects more effectively.

To infer the DM density from the kinematic data the Jeans analysis is used to relate the joint distribution of tracer stars' positions and velocities to the underlying potential of the system. Traditionally the second order Jeans equation \citep{binney} is used to generate the velocity dispersions for a set of input parameters including the potential which is then fitted to the dSph data with a likelihood analysis by radially binning the line of sight velocities for the variance and assuming Gaussianity. It has long been known however \citep{dejonghe87,merritt87} that this analysis may not be used to uniquely specify the potential for anisotropic systems for which the variances of the radial and tangential velocity components are not equal. As it is only possible to observe the projected quantities along the line of sight, the intrinsic dispersions of the system are convolved such that there is a degeneracy of indistinguishable solutions to the Jeans analysis. Indeed it has been shown that the observed line of sight dispersion can be generated by any given parameterisation of the anisotropy parameter \citep{evans2009} thus leaving the potential almost completely unconstrained.  This is the so-called Jeans degeneracy problem which is the main subject of this work.

A discussion of the higher order moments is presented herein with a mathematical description of how they enter the Jeans analysis and what assumptions are required to ensure that the Jeans degeneracy is solved or at least partially lifted. This is then placed into the practical context of a joint likelihood analysis of the variance and kurtosis in dwarf spheroidal galaxies. We evaluate the contribution by \cite{Lokas02} in establishing a model for the kurtosis that may be used to lift the degeneracy \citep{Lokas05} and extend the method to general anisotropy as proposed by \cite{an11a} with the separable augmented density system. 

To simplify the mathematical description of the higher order Jeans equations there has been much success in the literature since the advent of the augmented density formalism by \cite{dejonghe86}. Whilst an application \citep{dejonghe87,baes07} of this method has generally been limited to models \citep{plummer,hernquist} with particularly simple potential-density pairs, the recent work of An (2011b) demonstrates for generic density and anisotropy that a separable system \citep{Ciotti} solves the Jeans degeneracy problem completely by specifying moments at all orders with the potential and anisotropy parameter alone. This is however by no means a general solution and without a strong physical motivation its practical use is difficult to evaluate. An alternative hierarchy of \textit{pseudomoments} \citep{King} in spherical systems \citep{Amendt}, tailored to minimise the increasing dependence of standard moments to the tails of the distribution, breaks the degeneracy with physical arguments for weak nonisothermality. A key issue with the pseudomoment method is accessibility of the observable standard moments which has not yet been shown to be universal. With practical intent we thus persist with the standard moments for direct comparison with the data.  
 
Recently there have been a number of alternative original methodologies presented for breaking the degeneracy. \cite{penarrubia} and \cite{amoriscomulti} utilise the existence of chemodynamically distinct stellar populations in the Fornax and Sculptor galaxies to exploit the robustness of the mass profile to degenerate anisotropy at the stellar half radius and are able to derive an estimate of the mass slope $\Gamma = d\log M / d\log r$ that places stringent constraints on cusped density profiles. When taken at face value this observation, together with the additional apparent problem of missing satellites \citep{pen12}, is in tension with the $\Lambda$CDM model (although the role of baryons in the shaping the inner core of dark matter halos is very complicated \citep{governato}). Though powerful this method relies on multiple populations that may not exist in other dwarf spheroidals such as Carina \citep{penarrubia} and assumes Gaussianity in the line-of-sight velocity profiles that is inconsistent with the Jeans equations at orders higher than two. Whilst an application of the Jeans analysis to multiple populations is straightforward, at higher orders the number of input parameters quickly becomes impractical and more importantly still, splitting the population increases the errors associated with limited sampling. One way to mitigate this problem that is employed in the analysis of elliptical galaxies \citep[see e.g][]{Bender} is to use Gauss-Hermite moments \citep{Gerhard, Franx} that efficiently measure the shape of the distribution with less reference to the tails of the distribution than the traditional kurtosis. An extension of this method to discrete data sets \citep{Amorisco} enables an efficient extraction of non-Gaussian shape parameters suitable for an analysis of dwarf spheroidals. Though such an analysis is statistically preferable to conventional moment analysis it is difficult to ensure that the parametrised prior distributions are both physical and exhaustive. As hinted by \cite{Gerhard} it would be interesting to see whether a Jeans-like analysis to the Gauss-Hermite moments is viable which would ensure that the shape parameters are fitted to equilibrated systems. If information from non-Gaussianities can break the degeneracy then we choose as a simple first step to investigate what the well-established Jeans formalism can tell us for spherically symmetric systems such as dwarf spheroidal galaxies. Numerical methods such as the orbit-superposition algorithm \citep{Schwarzschild} have also recently been applied to dwarf spheroidal galaxies \citep[see for e.g][]{Breddels,Jardel} that make no assumption on the form of the anisotropy and guarantee physical distribution functions thus providing an interesting complement to the weaknesses of the traditional analytic methods described above.         

The layout of our paper is as follows.  In section \ref{sec2} we review the mathematics behind the Jeans equation and line of sight calculations.  To utilise the constraining power of the fourth order statistics we are thus motivated to provide the full analytical set of fourth order solutions which is achieved with the introduction of an analog to the Binney anisotropy parameter at fourth order. This is outlined in section \ref{sec3} wherein we show how to construct a generic model for the projected fourth order moment to be incorporated into a joint likelihood analysis, extending to full generality the over-constrained method employed by \cite{Lokas05}. In section \ref{sec4} we outline the joint likelihood analysis of dispersion and kurtosis that through the Jeans equations allows a fit of the density and anisotropy parameters to moments extracted from LOS velocity data. Section 5 sees the method tested for a set of simulated dwarf spheroidal data sets and as a proof of concept we contrast directly the performance of the traditional and joint analysis in constraining the anisotropy and crucially the density parameters. Finally we will make some concluding remarks and outline our future research program. 

\begin{section}{Preliminary}\label{sec2}
\subsection{Moments of the Distribution Function}
In the study of stellar systems, a 6-dimensional function $f$ is used to specify the distribution \citep{jean} of stars in position and velocity space. For a spherically symmetric system this is related to the underlying gravitational potential $\Phi(r)$ by the time-independent and collisionless Boltzmann equation \citep{merry},
\begin{center}
\begin{eqnarray}\label{boltz}
\frac{\partial f}{\partial t} &=& v_r \frac{\partial f}{\partial r} + \left( \frac{v^{2}_{\theta}+v^{2}_{\phi}}{r}-\frac{d\Phi}{dr}\right) \frac{\partial f}{\partial v_r} \nonumber\\ &+& \frac{1}{r}(v^{2}_{\phi}\cot \theta - v_rv_{\theta})\frac{\partial f}{\partial v_{\theta}}  \\ &-&  \frac{1}{r}(v_{\phi}v_r+v_{\phi}v_{\theta}\cot \theta )\frac{\partial f}{\partial v_{\phi}} \nonumber\\ &=& 0. \nonumber
 \end{eqnarray}
\end{center}
Multiplying \eqref{boltz} by ${v^{l}_{r}v^{m}_{\theta}v^{n}_{\phi}}$ and then integrating over all velocities restates it in terms of its \textit{true} velocity moments
\begin{equation}\label{moms}
\nu \overline{v^{2i}_r v^{2j}_{\theta} v^{2k}_{\phi}} = \int v^{2i}_r v^{2j}_{\theta} v^{2k}_{\phi} f(r,\textbf{v}) d^3v.
\end{equation}
where $\nu(r)$, as an effective zeroth moment that marginalises the distribution function in velocity space, is the local density of stars. 
Due to the spherical symmetry of the system it is trivial to show by averaging over the azimuthal angles that the odd moments vanish and that many of the true even moments are related by constant prefactors. To make the notation more compact we again follow the example of \cite{merry} and introduce the \textit{intrinsic} moments of the tangential velocity $v_t=(v^{2}_{\theta}+v^{2}_{\phi})^{1/2}$, 
\begin{eqnarray}
\overline{v^{2i}_r v^{2j}_{\theta} v^{2k}_{\phi}} = \frac{1}{\pi} B(j+\frac{1}{2},k+\frac{1}{2})\overline{v^{2i}_r\; v^{2(j+k)}_t}
\end{eqnarray}
where $B(x,y)$ is the Beta function. In the subsequent analysis we find that it is not necessary to explicitly refer to the true moments at any stage and to simplify the mathematics the tangential moments will be used exclusively from here on in. 
\subsection{Jeans Equations}
To isolate the second order moments i.e the radial and tangential dispersions which in practice have the smallest statistical errors, the Boltzmann equation is traditionally multiplied by $v_r$ and integrated over all velocities \citep{binney} to give,
\begin{equation}\label{jeans}
\frac{d(\nu \sigma^{2}_{r})}{dr} + \frac{2\beta}{r}\nu \sigma^{2}_{r} +\nu \frac{d \Phi}{dr} = 0.
\end{equation}
where $\nu(r)$ is the local stellar density, $\Phi(r)$ is the gravitational potential that depends on the total density of the system $\rho(r) = \nu(r)+\rho_{\text{dm}}(r)$ via,
\begin{equation}\label{phi}
\Phi(r) = \frac{4\pi G}{r} \int^{r}_{0} r^2 \rho(r)dr
\end{equation}
and the Binney anisotropy parameter \citep{binney} $\beta(r)$,
\begin{equation}
\label{anis}
\beta(r) \equiv 1-\frac{\sigma^{2}_{t}(r)}{2\sigma^{2}_{r}(r)},
\end{equation}
measures the deviation of the dispersions from the isotropic system $(\sigma^{2}_{r}=\sigma^{2}_{\theta}=\frac{1}{2}\sigma^{2}_{t})$ wherein all directions in velocity space are equally probable\footnote{We choose for mathematical convenience to adopt the 2D tangential dispersion rather than the 1D employed by e.g \cite{Lokas02} which accounts for the additional factor of 2.}. For a dSph, where the mass-luminosity ratios are often greater than 10 \citep{mateo} the dark matter component is very significant.

To illustrate the higher order analysis we consider first the fourth order where multiplying equation \eqref{boltz} by $v^{3}_{r}$ and $v_r v^{2}_{\theta}$ respectively relates the three intrinsic moments at fourth order $\overline{v^{4}_{r}},\overline{v^{4}_{t}}$ and $\overline{v^{2}_{r}v^{2}_{t}}$ by the two fourth order Jeans equations \citep{merry},
\begin{equation}
\label{hojeans1}
\frac{d(\nu \overline{v^{4}_{r}})}{dr} - \frac{3}{r}\nu \overline{v^{2}_{r}v^{2}_{t}}  + \frac{2}{r}\nu \overline{v^{4}_{r}}  + 3 \nu \sigma^{2}_{r} \frac{d \Phi}{dr} = 0
\end{equation}
\begin{equation}
\label{hojeans2}
\frac{d(\nu \overline{v^{2}_{r}v^{2}_{t}})}{dr} - \frac{1}{r}\nu \overline{v^{4}_{t}}  + \frac{4}{r}\nu \overline{v^{2}_{r}v^{2}_{t}}  +  \nu \sigma^{2}_{t} \frac{d \Phi}{dr} = 0.
\end{equation}
The advent of the augmented density system by \cite{dejonghe86} has greatly enhanced the mathematical description of the Jeans analysis and it is within this framework that the complete set of Jeans equations has been presented \citep{an11b}, 
\begin{eqnarray}\label{genjean}
\frac{d(\nu \overline{v^{2p}_{r}v^{2q}_{t}})}{dr} &=& -\frac{2}{r}\left[(q+1)\nu \overline{v^{2p}_{r}v^{2q}_{t}}-(p-\frac{1}{2})\nu \overline{v^{2p-2}_{r}v^{2q+2}_{t}}\right] \nonumber\\ &&-(2p-1)\nu \overline{v^{2p-2}_{r}v^{2q}_{t}}\frac{d\Phi}{dr}. 
\end{eqnarray}
The number of equations at 2$n$th order is therefore $n$, the number of permutations of $(p,q)$ for which $p+q=n$ and $1\leq p \leq n,\; 0 \leq q \leq n$. Each of the $n$ moments at 2$n$th order enter the derivative of a corresponding equation apart from $v^{2n}_{t}$.

We also note that as the distribution function is linear upon disassembling into $M$ stellar sub-components, it follows from \eqref{moms} that the composite moments are related to the constituent moments via
\begin{equation}
\nu \overline{v^{2i}_{r}\; v^{2j}_{t}}=\sum^{M}_{c=1}\nu_c (\overline{v^{2i}_r\; v^{2j}_t})_c
\end{equation}
where $\nu_c$ is the constituent stellar density that intuitively weights the relative contribution from each sub-population. One can then show that the Jeans equations are also linear and thus that solving a Jeans equation for a  composite distribution function is equivalent to simultaneously solving the sum of equations for each individual population. Indeed this must be true for stars orbiting in a common potential (like the dark matter dominated potential of a dwarf spheroidal galaxy) else one could not arbitrarily select sub-samples from the data that were equilibrated. For chemically distinct populations with different histories however we reemphasise this point and retain the larger composite population to reduce the sampling errors that are critical in a higher moment analysis.    
    
\subsection{Projected Moments}
Unfortunately due to the distant nature of astronomical objects the stellar positions and velocities may only be observed along the line-of sight and rather than the true moments and the local density one must instead use the \textit{projected} moments and the surface density profile $ \Sigma $ as functions of the projected radius $R$.
The surface density profile is the projection along the line of sight of $\nu(r)$,
\begin{equation}
\Sigma(R) = 2\int^{\infty}_{R} \frac{\nu(r)r}{\sqrt{r^2-R^2}}dr.
\end{equation}
which may be inverted directly via the Abel Inversion for the local density. The component of the velocity along the line of sight, which for convenience is chosen as the z-direction, may be expressed as 
\begin{equation}\label{losz}
v_{\text{los}} = v_r \cos(a) - v_{\theta} \sin(a) = v_r \sqrt{1-\frac{R^2}{r^2}} - v_{\theta}\frac{R}{r}.
\end{equation}
where $\sin a = R/r$ is the unobservable depth angle that determines the extent to which each stellar velocity is radial or tangential. Finding the projected moment at 2$n$th order \citep[see also][]{dejonghe92} is akin to averaging over the 2$n$th power of equation \eqref{losz},
\begin{eqnarray}\label{nlosgen}
\Sigma\overline{v^{2n}_{\text{los}}}(R) &=&  2\int^{\infty}_{R} \overline{(v_r \cos a-v_{\theta}\sin a)^{2n}} \frac{\nu(r) r}{\sqrt{r^2-R^2}}dr \nonumber\\
&=& 2\int^{\infty}_{R} \sum^{n}_{k=0} C_{n,k} \overline{v^{2(n-k)}_{r}v^{2k}_{t}}  \frac{\nu(r) r}{\sqrt{r^2-R^2}}dr
\end{eqnarray}
and one thus requires all of the intrinsic moments to calculate the projection with the coefficients,
\begin{equation}
C_{n,k} = \binom{2n}{2k}\frac{B(k+\frac{1}{2},\frac{1}{2})}{\pi} \left(1-\frac{R^2}{r^2}\right)^{n-k}\left(\frac{R^2}{r^2}\right)^{k}.
\end{equation}
To simplify the projected dispersion the anisotropy parameter is again introduced in place of the tangential dispersion,
\begin{equation}\label{LOSsecond}
\Sigma \sigma^{2}_{\text{los}}(R) = 2\int^{\infty}_{R} (1-\beta\frac{R^2}{r^2}) \frac{\nu \sigma^{2}_{r} r}{\sqrt{r^2-R^2}}dr 
\end{equation}
such that specifying the potential and anisotropy parameter is sufficient to first calculate the radial dispersion via equation \eqref{jeans} and then the projected moment for comparison with observation. An explicit calculation of the projected fourth moment
\begin{equation}
\label{LOSfour}
\Sigma\overline{v^{4}_{\text{los}}}(R) =  2\int^{\infty}_{R} \left(C_{2,0} \overline{v^{4}_{r}} + C_{2,1}\overline{v^{2}_{r}v^{2}_{t}} +C_{2,2}\overline{v^4_t}\right) \frac{\nu(r)r}{\sqrt{r^2-R^2}}dr
\end{equation}
\begin{equation}
C_{2,k} = \left\{ \begin{array}{cc} (1-\frac{R^2}{r^2})^2 & k=0\\ 3 \frac{R^2}{r^4}(r^2-R^2) & k=1\\ \frac{3}{8} \frac{R^4}{r^4} & k=2 \end{array}\right.
\end{equation}
recovers the result in \cite{merry}.
\subsection{The Degeneracy Problem}
The normal Jeans analysis, which approximates the distribution function by its second order moments, has traditionally been the only viable means of modeling the limited sample sizes afforded by kinematic surveys of dwarf spheroidal galaxies. Unfortunately it has been demonstrated \citep[e.g][]{merritt87} that the integral equation \eqref{LOSsecond} can be highly degenerate with no way of distinguishing the entangled intrinsic dispersions. In a typical statistical treatment, where a model for the dark matter density is fitted with a hand-picked form of the anisotropy, there are numerate parameter sets $p=\{\beta(r),\Phi(r)\}$ that yield identical line of sight dispersions within statistical errors. As the anisotropy parameter is a completely unknown degree of freedom it is impossible to uniquely specify the potential with the line-of-sight dispersion alone and such a treatment is liable not only to imprecision but also to inaccuracy.
With a recent improvement in the data, there has been a renewed interest in the higher order moments that may be used to distinguish those parameter sets degenerate at second order. Unfortunately whilst the fourth moment is practically within reach, a theoretical problem persists. As suggested by \cite{an11b} we note that at each successive order the Jeans analysis introduces $n+1$ moments and only $n$ constraining Jeans equations such that the intrinsic moments are not specified by the second order parameters $p$. A minimal requirement to define the system at fourth order is to specify one of the fourth order moments or, in analogy with Binney's anisotropy parameter, to specify the ratio of two. To lift the degeneracy one additionally desires that the projected fourth moment depend only upon the second order parameters such that a new net constraint is added to the system without expanding the parameter space. The anisotropy parameter in particular, whilst inherent to the dispersions, has no direct bearing on the higher order moments without artificial insertion. It can be concluded therefore that to utilise the higher order moments one must present a new constraint to the system via a simplifying assumption or empirical observation that optimally ties the higher order moments to the anisotropy parameter and thus may be used to constrain it. To lift the degeneracy completely one requires all of the projected moments which as proved by \cite{dejonghe92} is equivalent to knowledge of the distribution function.
\subsection{Incompatibility of Equilibrium and the Assumption of Gaussian Velocity Distributions}
Whilst an additional constraint is required for unique solutions to the fourth order Jeans equations, imposing more than one over constrains the equations such that there is no consistent equilibrium solution. If one assumes that the joint distribution of radial and the tangential velocities is normal and separable,
\begin{equation}
f(v_r,v_\theta,v_\phi) =  \frac{1}{(2\pi)^{3/2} \sigma^{2}_{\theta} \sigma_r }\exp\left(-\frac{v^{2}_{r}}{2\sigma^{2}_{r}}-\frac{v^{2}_{\theta}+v^{2}_{\phi}}{2\sigma^{2}_{\theta}}\right)
\end{equation}
then all higher orders are fixed. At fourth order the moments follow from \eqref{boltz}
\begin{equation}\label{gmoms}
\overline{v^{4}_{r}} = 3 \sigma^{4}_{r}
\end{equation} 
\begin{equation}
\overline{v^{2}_{r}v^{2}_{t}} = 2\overline{v^{2}_{r}v^{2}_{\theta}} = 2\sigma^{2}_{r}\sigma^{2}_{\theta} = (1-\beta)\sigma^{4}_{r}
\end{equation}
\begin{equation}
\overline{v^{4}_{t}} = \frac{8}{3}\overline{v^{4}_{\theta}} = 8 \sigma^{4}_{\theta}= \frac{1}{2}(1-\beta)^2\sigma^{4}_{r}.
\end{equation}
imposing three constraints to the system. Let's now take a step back and say that we enforce only the first, that the radial velocity distribution is Gaussian. From \eqref{hojeans1} there is a unique solution for the comoment $\overline{v^{2}_{r}v^{2}_{t}}$ where the RHS is zero. This in turn yields $\overline{v^{4}_{t}}$ from \eqref{hojeans2} and thus the one additional constraint specifies a unique equilibrated system. Introducing the other two constraints however completely specifies the LHS of the Jeans equations which is not guaranteed to be (in fact it is exceptionally rarely) zero. Recalling that to generate \eqref{hojeans1}, the CBE \eqref{boltz} is multiplied through by $v^{3}_{r}$ and integrated over all velocities we perform the same to the time derivative which yields, 
\begin{equation}
\int v^{3}_{r} \frac{\partial f}{\partial t} d^3 v = \frac{d \nu \overline{v^{3}_{r}}}{dt}
\end{equation}
i.e the rate of change (when normalised by the dispersions) of skewness in the radial velocity distribution. With the Gaussian assumption then plugging \eqref{gmoms} into the fourth order Jeans equations enables a calculation of this effect for a given set of parameters,
\begin{equation}
\frac{d\nu \overline{v^{3}_{r}}}{dt} =  3\nu\sigma^{2}_{r}\left[\left(\frac{1-3\beta}{r}-\frac{1}{\nu}\frac{d\nu}{dr}\right)\sigma^{2}_{r}-\frac{d\Phi}{dr}\right].
\end{equation}      
\end{section}
\section{Analytical Solutions to the Jeans Degeneracy}\label{sec3}
In this section we discuss analytical models that place additional constraints on the intrinsic fourth order moments for a practical use in tackling the mass-anisotropy degeneracy. One prominent example in the literature is the work of \cite{Lokas02} who assumes a form \citep{henon} of the distribution function $f(E,L) = f_0(E)L^{-2\beta}$ with a constant velocity anisotropy that is separable when expressed in terms of the specific binding energy $E = -\Phi - \frac{1}{2}(v^{2}_{r}+ v^{2}_{t})$ and the specific angular momentum $|L|=rv_t$. With this assumption then without specifying the form of the energy component $f_0(E)$ it is possible via equation \eqref{moms} to calculate the ratio of higher order moments with the anisotropy parameter. The projected fourth moment is thus shown to be dependent only on the second order parameters and is applied to the degeneracy problem for the Draco galaxy \citep{Lokas05}. It has been argued however \citep{an11b} that this model, whilst relatively clear to interpret, over-constrains the problem and may be extended to general anisotropy with the separable augmented density model. As such we outline this model and provide a formula for the projected moments. This system however only provides one solution and whilst mathematically elegant it does not yet have a strong physical motivation. We therefore provide a framework without appealing to the augmented density formalism directly, that encompasses all solutions to the fourth order by introducing an analog to the anisotropy parameter at fourth order. The aim of this framework is both to evaluate the particularly convenient separable augmented solution and to facilitate more physically motivated models perhaps inspired by empirical observation. The framework is then extended to all orders. 

\subsection{The Separable Augmented Density Model}
A full account of the augmented density formalism is not necessary for the wider context of this paper but we direct the interested reader to the literature \citep{dejonghe86,dejonghe92,an11b,an11a}. Here we give a brief outline and then present the main findings relevant to a study of dwarf spheroidal galaxies in the simplest possible terms.
By careful consideration of equation \eqref{moms} and the Jeans theorem \citep{jean} it was noted by \cite{dejonghe86} that the degeneracy of distribution functions which fit the local density $\nu(r)$ can be represented by \textit{augmenting} the density into an infinite set of degenerate bi-variate functions of the \textit{augmented configuration space} $(\Psi,r)$ and that providing a specific form of the augmented density is equivalent to providing the distribution function. Here $\psi$ is the positive binding specific potential defined by $\frac{d\Psi}{dr}=-\frac{d\Phi}{dr}$. Moments are also augmented and to retrieve the observables the augmented quantities are \textit{deaugmented} by taking the limit $\Psi \rightarrow \Psi(r)$. The power of the method lies in the ease with which one may relate the augmented quantities and in augmented configuration space the Jeans equations are replaced by the simple partial differential equations \citep{dejonghe92}, 
\begin{equation}\label{psider}
\frac{\partial}{\partial \Psi} \nu \overline{v^{2p}_{r}v^{2q}_{t}}(\Psi,r) = (2p-1) \nu \overline{v^{2p-2}_{r}v^{2q}_{t}}(\Psi,r)
\end{equation}
\begin{equation}\label{rder}
\frac{\partial}{\partial r^2}\left[r^{2q+2}\nu \overline{v^{2p}_{r}v^{2q}_{t}}(\Psi,r)\right] = (p-\frac{1}{2})r^{2q}\nu \overline{v^{2p-2}_{r}v^{2q+2}_{t}}(\Psi,r)
\end{equation}
and the augmented total derivative \cite{an11b},
\begin{equation}
\frac{d}{dr}\rightarrow\frac{\partial}{\partial r} + \frac{\partial}{\partial \Psi}\frac{d\Psi}{dr}.
\end{equation}
To determine the moment ratios at fixed order it is necessary only to consider the order preserving equation \eqref{rder} such that for a 
system with a separable augmented density,
\begin{equation}\label{sepaug}
\widetilde{\nu}(\Psi,r) = P(\Psi)R(r)
\end{equation}
the anisotropy is dependent only on the radial component $R(r)$ and this has been demonstrated by \cite{Ciotti} who shows that
\begin{equation}
\beta = - \frac{d \log R}{d \log r^2}.
\end{equation}
An even more profound property of the separable system (equation \ref{sepaug}) is that the ratio of moments at a given order scales only with the constant prefactor inherent to the isotropic system. In practice this has the notable effect of collapsing all of the Jeans equations to just one \citep{an11b},
\begin{equation}\label{njeans}
\frac{d(\nu \overline{v^{2n}_{r}})}{dr} + \frac{2\beta}{r} \nu \overline{v^{2n}_{r}} + (2n-1)\nu \overline{v^{2n-2}_{r}}\frac{d\Phi}{dr} = 0
\end{equation}
that enables a unique calculation of the radial intrinsic moment and thus absorbs the additional degree of freedom in the Jeans analysis. In calculation of the other moments one may use \citep{an11b},
\begin{equation}
\nu \overline{v^{2(n-k)}_{r}v^{2k}_{t}} = \frac{\alpha_k}{(n-k+\frac{1}{2})_k} \nu \overline{v^{2n}_{r}}
\end{equation}
where $(a)_k = \prod^{k}_{i=1}(a+i-1)$ is the Pochhammer symbol that describes the $\textit{rising sequential product}$, $\alpha_0=1$ and
\begin{equation}\label{Anal}
\alpha_{q+1} = (q+1-\beta)\alpha_{q}+\frac{r}{2}\frac{d\alpha_q}{dr}
\end{equation}
can be used to iteratively generate all moment ratios noting again that one requires only the anisotropy parameter. The projected moments then follow trivially,
\begin{equation}\label{nlossep}
\Sigma\overline{v^{2n}_{\text{los}}}(R) = 2\int^{\infty}_{R}\sum^{n}_{k=0}C_{n,k}\frac{\alpha_{k}}{(n-k+\frac{1}{2})_{k}} \frac{\nu \overline{v^{2n}_{r}}r}{\sqrt{r^2-R^2}}dr
\end{equation}
and we thus demonstrate that the fourth projected moment depends only on the second order parameters. In practice one must first solve $n$ differential equations to calculate in turn each radial intrinsic moment via the Jeans equation \eqref{njeans} and then $n$ iterations of \eqref{Anal} for a theoretical prediction of the projected moment.   
As an example consider the projected fourth moment for which we require,
\begin{equation}\label{alfour}
\alpha_1 = 1-\beta,\;\;\;\;\;\alpha_2=(1-\beta)(2-\beta)-\frac{r}{2}\frac{d\beta}{dr}.
\end{equation}
and the fourth order Jeans equation for a separable augmented density,
\begin{equation}\label{fourjeans}
\frac{d\nu\overline{v^{4}_{r}}}{dr} + \frac{2\beta}{r}\nu\overline{v^{4}_{r}} + 3\nu\sigma^{2}_{r}\frac{d\Phi}{dr}=0.
\end{equation}
Evaluating \eqref{nlossep} with $n=2$ yields after some algebra,
\begin{equation}\label{LOSfoursep}
\Sigma\overline{v^{4}_{los}}(R) =  2\int^{\infty}_{R} g(\beta,r,R) \frac{\nu\overline{v^{4}_{r}} r}{\sqrt{r^2-R^2}}dr
\end{equation}
\begin{equation}\label{gfunc}
 g(\beta,r,R)=1-2\beta\frac{R^2}{r^2}+\frac{\beta(1+\beta)}{2}\frac{R^4}{r^4}-\frac{R^4}{4r^3}\frac{d\beta}{dr} 
\end{equation}
which generalises the result of \cite{Lokas02} and may be used to tackle the Jeans degeneracy for an arbitrary radial dependence of the anisotropy parameter via the method employed in \cite{Lokas05}. In practice then one must first find the radial dispersion via the second order Jeans equation \eqref{jeans}, solve equation \eqref{fourjeans} for the fourth radial moment and then with the recursive relations \eqref{alfour} it is possible to evaluate the projected moment with equation \eqref{LOSfour}.

To illustrate the limitations of the separable augmented density model in describing the anisotropy between the 1D radial velocity distribution $P_r(v_r)$ and its 1D tangential counterpart $P_\theta(v_\theta)=P_\phi(v_\phi)$ we determine the kurtosis of the 1D tangential distribution which for constant anisotropy $\beta$ is
\begin{equation}\label{lokaskurt}
\kappa_\theta=\frac{3\overline{v^{4}_{t}}}{2 \sigma^{4}_{t}} = \frac{2-\beta}{2(1-\beta)}\kappa_r
\end{equation}
where  $\kappa_r = \overline{v^{4}_{r}}/\sigma^{4}_{r}$ is the kurtosis of $P(v_r)$.
\begin{figure}
	\centering
		\includegraphics[width=9.5cm]{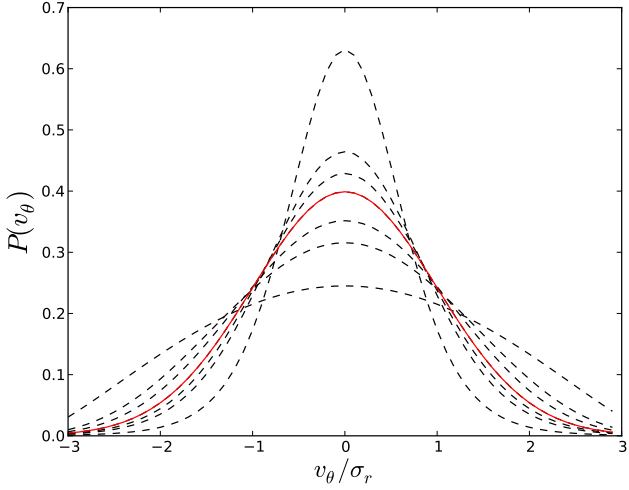}
	
	\caption{Anisotropy in the Henon-Lokas model: Assuming a Gaussian distribution for $P_r$ plotted in red we plot $P_\theta$ for $\beta$ in the range [-1,0.5] assuming a Pearson distribution (see Appendix B) with variance $\sigma^{2}_{\theta}=(1-\beta)\sigma^{2}_{r}$ and kurtosis from \eqref{lokaskurt}. }
	\label{lokasdists}
\end{figure}
 As illustrated in Fig.\ref{lokasdists} this limits the possible $P_\theta$ distributions such that those with greater width than $P_r$ must also have heavier tails. If one fixes the widths of the distributions to be equal then one is left with only the isotropic solution and no possible anisotropy in other shape parameters.   

In summary the separable augmented density system completely lifts the Jeans degeneracy by providing an infinite set of projected moment equations that one can, in principle, calculate from the potential and the anisotropy at second order. The origin of the system is however purely mathematical and whilst particularly convenient to use it is not unique in exhibiting this behaviour. Indeed one could artificially introduce any arbitrary relationship between the anisotropy parameter and the higher order moments to the same effect. This model, which generates only one DF for a given set of anisotropy and density parameters, is thus a severe restriction to apply to the system without physical motivation. The aim then is to provide a framework that explores the full range of solutions to the Jeans degeneracy problem without restriction from mathematical considerations.
\subsection{An Extended Model of Anisotropy}
In this subsection we will demonstrate, without appealing explicitly to the augmented density formalism, that a unique set of of all deprojected moments is specified by introducing an analog to the Binney anisotropy parameter at each successive order. Indeed one can represent all possible distribution functions in this way and we will show that a distribution function can be defined as an infinite set of anisotropy parameter analogs $\{\beta_n\}$, $1\leq n \leq \infty$. If each of these is assumed to have a known relation to the anisotropy parameter then the Jeans degeneracy is completely lifted. For a particular definition of anisotropy parameters, inspired by the separable augmented density system, we provide formulae for the full set of intrinsic moments and their subsequent projected moments. We are then able to interpret the separable augmented density system as a particular subset of this system for which the anisotropy parameter analogs are independent of order $\{\beta_n\}=\beta\;\forall n$. 
\subsubsection{Fourth Order}
\begin{figure}
	\centering
		\includegraphics[width=9.5cm]{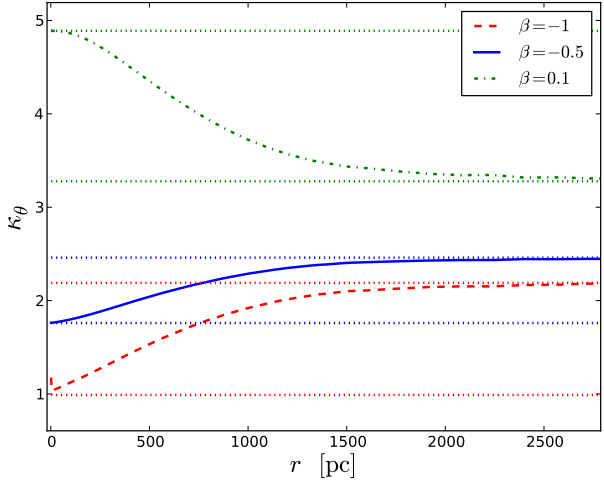}
	
	\caption{For a galaxy with stars following a Plummer profile with radius$ r_p=575pc$, an Einasto DM profile (see Section 4) with parameters $\{\rho_{-2}=0.05 M_{\rm{sol}}\rm{pc}^{-3},\; r_{-2}=700\rm{pc},\;\alpha=2\}$ and constant $\beta^{\prime}=-0.4$ then assuming that the distribution of deprojected radial velocities is Gaussian ($\kappa_r = 3$) the kurtosis \eqref{kurtthet} is plotted for the constant anisotropy parameters shown with approximations to the asymptotic limits at $r\to 0$ \eqref{kzero} and $r\to \infty$ \eqref{kinfty} plotted as dotted lines.} 
\label{kurtradprof}
\end{figure}

To measure the anisotropy at fourth order we choose for simplicity the adjacent moment ratio to the radial fourth moment and introduce,
\begin{equation}\label{betprime}
\beta'(r) = 1-\frac{3}{2}\frac{\overline{v^{2}_{r}v^{2}_{t}}}{\overline{v^{4}_{r}}},
\end{equation}
which is an analog of the Binney anisotropy parameter for the fourth order that measures the deviation from the isotropic system \eqref{prodad} where ${\overline{v^{4}_{r}}} = \frac{3}{2}\overline{v^{2}_{r}v^{2}_{t}}$. Substituting this parameter into the Jeans equation \eqref{hojeans1}, 
\begin{equation}\label{radeq}
\frac{d(\nu \overline{v^{4}_{r}})}{dr} + \frac{2 \beta'}{r}\nu \overline{v^{4}_{r}} + 3 \nu \sigma^{2}_{r} \frac{d \Phi}{dr} = 0,
\end{equation}
which then, given a functional form for $\beta'$, enables the radial fourth moment to be determined uniquely. The choice of equation \eqref{betprime} is now clear as we note that the separable augmented density system thus corresponds to $\beta=\beta'$. The mixed moment then follows trivially from the definition of $\beta'$ and the tangential moment may be calculated via equation \eqref{hojeans2} as follows. Firstly equation \eqref{betprime} indicates that,
\begin{equation}
\frac{d(\nu \overline{v^{2}_{r}v^{2}_{t}})}{dr} = \frac{2}{3}\left[(1-\beta') \frac{d(\nu \overline{v^{4}_{r}})}{dr} - \frac{d\beta'}{dr}\nu  \overline{v^{4}_{r}}\right]
\end{equation}
which may be simplified further with equation \eqref{radeq} by substituting in the derivative of the radial fourth moment,
\begin{equation}
\frac{d(\nu \overline{v^{2}_{r}v^{2}_{t}})}{dr} = \frac{2}{3}(1-\beta')\left[-\frac{2 \beta'}{r} \nu \overline{v^{4}_{r}} - 3 \nu \sigma^{2}_{r} \frac{d \Phi}{dr}\right]  - \frac{2}{3} \frac{d\beta'}{dr}\nu  \overline{v^{4}_{r}}.
\end{equation}
Plugging this into the Jeans equation \eqref{hojeans2} and rearranging for the fourth tangential moment yields,
\begin{equation}\label{tanfourgen}
\overline{v^4_t} = \frac{4}{3}\left((1-\beta')(2-\beta') - \frac{r}{2} \frac{d\beta'}{dr}\right) \overline{v^4_r}+2(\beta'-\beta)r\sigma^{2}_{r}\frac{d\Phi}{dr}
\end{equation}
which shows that the system is uniquely determined by the second order parameters plus the anisotropy at fourth order. One also notes that in the limit $\beta'=\beta$, the tangential moment recovers the result $\overline{v^4_t} = \frac{4}{3}\alpha_2 \overline{v^4_r}$ for the separable augmented density system. With this result it is possible to gain a better understanding of the effect that $\beta^{\prime}$ has on the model. Where $\beta$ indicates directly the relative width of the 1D velocity distributions $P_r(v_r)$ and $P_{\theta}(v_{\theta})$, the primary effect of $\beta^{\prime}$ on the relative kurtosis is not so easy to interpret. From \eqref{tanfourgen} we derive the relationship between kurtosis for this generalised model,  
\begin{equation}\label{kurtthet}
 \kappa_{\theta} = \frac{(1-\beta')(2-\beta') - \frac{r}{2} \frac{d\beta'}{dr}}{2(1-\beta)^2}\kappa_r + \frac{3r(\beta'-\beta)}{4\left(1-\beta\right)^{2}\sigma^{2}_{r}}\frac{d\Phi}{dr}
\end{equation}
where the new correction to the separable augmented density depends not only on the discrepancy $\beta^{\prime}-\beta$ but also on the form of the potential. At large radii where this correction is often dominant, the dependence on the radial dispersion requires a numerical calculation of the second order Jeans equation. For an analytic form we consider the limit as $r\to\infty$ which is constant under certain assumptions for the parameterisation of the stellar density and anisotropy profiles which we employ in section 4, namely a Plummer profile for the stellar density $v(r)$ with $ d\ln \nu/dr \to \frac{5}{r}$, a dark matter density profile  (such as the Einasto profile) that asymptotes to a constant enclosed mass $ M(r) \to M_\infty$ and a parameterisation of the anisotropy parameters that asymptotes to a constant $\beta^{(\prime)}_\infty$ with vanishing derivative $ d\beta/dr \to 0$. Under these assumptions the Jeans equation is straightforward to evaluate with
\begin{equation}
\lim_{r\to \infty} \sigma^{2}_{r} = \frac{GM_\infty}{2(3-\beta_\infty)r}.
\end{equation}
which yields a limit that is independent of the total enclosed mass $M_\infty$,
\begin{equation}\label{kinfty}
\lim_{r\to \infty} \kappa_\theta = \frac{(1-\beta^{\prime}_{\infty})(2-\beta^{\prime}_{\infty})}{2(1-\beta_\infty)^2}\kappa_r + \frac{3}{2}\frac{(\beta^{\prime}_{\infty}-\beta_\infty)(3-\beta_\infty)}{(1-\beta_\infty)^2}
\end{equation}
Assuming that the anisotropy asymptotes to a constant value $\beta_0$ as $r\to0$ then for the stellar and DM densities considered in section 4 with constant density cores $M(r) \propto r^3$ the beta term dominates the Jeans equation and we find,
\begin{equation}
\lim_{r\to 0} \sigma^{2}_{r} = r^{-2\beta_0}
\end{equation}
such that the corrective term vanishes at least linearly for $\beta>-\frac{1}{2}$ and we are left with,
\begin{equation}\label{kzero}
\lim_{r\to 0} \kappa_\theta = \frac{(1-\beta^{\prime}_{0})(2-\beta^{\prime}_{0})}{2(1-\beta_0)^2}\kappa_r \;,\;\;\beta_0>-\frac{1}{2}
\end{equation}
 In Fig. \ref{kurtradprof} we show examples of the radial profile as calculated numerically from the Jeans equation. The curves shown, which have constant anisotropy and density profiles well described by the assumptions listed above, vary smoothly between the asymptotic limits. For the curve with a high degree of tangential anisotropy we see that the approximation \eqref{kzero} breaks down and if $\beta$ was decreased further we would see a larger and larger discrepancy with $\kappa_\theta$ shooting up to infinity at the centre of the galaxy.
\begin{figure}
	\centering
		\includegraphics[width=9.5cm]{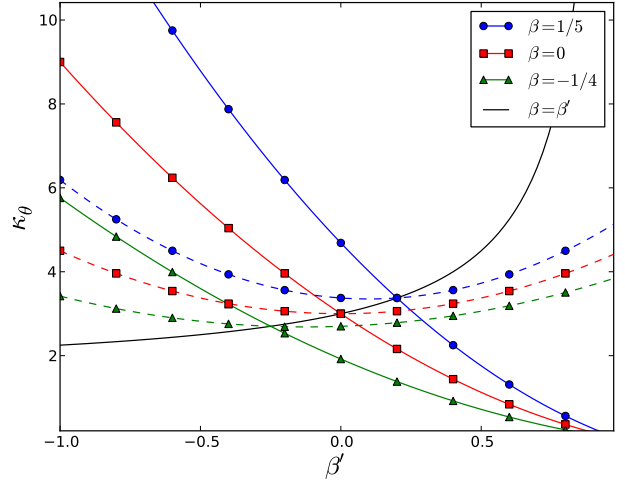}
	
	\caption{Anisotropy in the generalised model: Assuming that $P_r$ is Gaussian the kurtosis of $P_\theta$ at all radii is approximately enclosed in the region between the asymptotic values at $r=0$ (solid) and $r\to\infty$ (dashed). For reference we show the Henon-Lokas contour $\beta=\beta^{\prime}$ that corresponds to the 1D radial and tangential distributions displayed in Fig.\ref{lokasdists}. }
	\label{betprimekurts}
\end{figure}
\begin{figure}
	\centering
		\includegraphics[width=9.5cm]{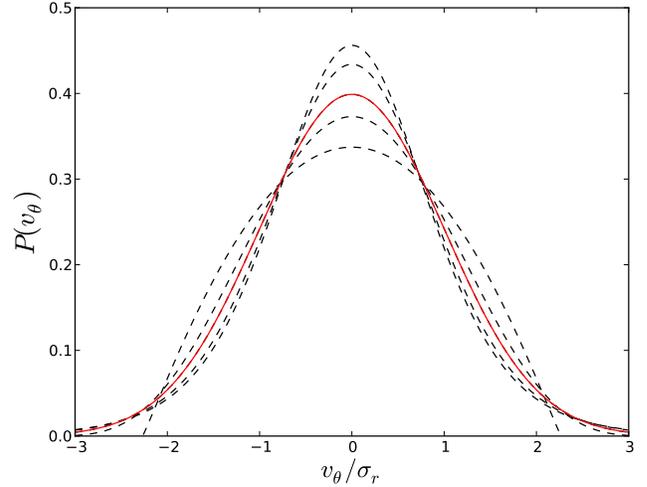}
	
	\caption{Affect of $\beta^{\prime}$ on anisotropy in the generalised model: Again assuming $P_r$ to be a Gaussian with width $\sigma_r$ (red), the 1D tangential distribution $P_\theta$ at $r=0$ is plotted for fixed $\beta=0$ (and thus fixed variance $\sigma_\theta = \sigma_r$) for $\beta^{\prime}$ in the range [-0.4, 0.2] assuming a Pearson distribution (Appendix B) with kurtosis derived from \eqref{kzero}.}
	\label{kurtzero}
\end{figure}
Fig. \ref{betprimekurts} shows the additional freedom afforded by $\beta^{\prime}$ to describe all kurtosis values regardless of $\beta$ at small radii though in the assumption of constant anisotropy parameters there is a convergence at larger distances. Positive deviations $\beta^{\prime}>\beta$ give relatively flat topped distributions compared to the Henon-Lokas model and thus as $\beta>0$ describes radially biased widths, then if one holds $\beta=0$, the condition $\beta^{\prime}>0$ gives radially biased kurtosis as shown in Fig. \ref{kurtzero}.   
 With all the moments one can then compute the projected moment from equation \eqref{LOSfour} which after some algebra is,  
\begin{equation}\label{LOSgenfour}
\Sigma\overline{v^{4}_{los}}(R) =  2\int^{\infty}_{R} \left(g(\beta')\overline{v^4_{r}}+\frac{3R^4}{4r^3}(\beta'-\beta)\sigma^{2}_{r}\frac{d\Phi}{dr}\right) \frac{\nu(r)r}{\sqrt{r^2-R^2}}dr
\end{equation}
where $g(\beta')$ is adapted from equation \eqref{gfunc} with $\beta\rightarrow\beta'$. If one assumes constant anisotropy at both orders then one may use the simple integral form \citep{Lokas05},
\begin{equation}
\nu \sigma^{2}_{r}(\beta=\rm{const})=r^{-2\beta}\int^{\infty}_{r}r^{2\beta}\nu\frac{d\Phi}{dr}dr
\end{equation} 
\begin{equation}
\nu \overline{v^{4}_{r}}(\beta^{\prime} =\rm{const})=3r^{-2\beta^{\prime}}\int^{\infty}_{r}r^{2\beta^{\prime}}\nu\sigma^{2}_{r}\frac{d\Phi}{dr}dr
\end{equation} 
and thus from an analytical standpoint the extension of Lokas's model is straightforward to implement.

 An arbitrary choice of $\beta'(r)$ in addition to the second order parameters represents the complete set of fourth order systems. Therefore to lift the Jeans degeneracy one must additionally impose that $\beta' = f(\beta)$ where a specification of the function $f(\beta)$ defines the model. This is the fundamental concept of this paper. In section \ref{sec4} we explore the effect of this choice on the key physical observables of the dark matter problem. If one can use empirical evidence to find a correlation between the anisotropy at both orders or indeed a well motivated physical argument for the specific form that such a relationship would take, then the Jeans degeneracy problem can be at least partially lifted with available data. To lift the degeneracy completely, all projected moments are required and in the following section the framework presented above is extended.
\subsubsection{Anisotropy of the Entire System}
To generalise the method in the previous section we again choose to define the anisotropy parameter analog at 2$n$th order via the adjacent moment ratio to the radial moment,
\begin{equation}
\beta_n(r) = 1-(n-\frac{1}{2})\frac{\overline{v^{2n-2}_{r}v^{2}_{t}}}{\overline{v^{2n}_{r}}}
\end{equation}
which measures the departure from the isotropic system, $\overline{v^{2n}_{r}} = (n-1/2)\overline{v^{2n-2}_{r}v^{2}_{t}}$ and we note that Binney's anisotropy parameter $\beta = \beta_1$ and its fourth order analog $\beta' = \beta_2$ are naturally incorporated into the analysis. Substituting this into the first Jeans equation at $2n$th order,
\begin{equation}\label{genjeans}
\frac{d(\nu \overline{v^{2n}_{r}})}{dr} + \frac{2\beta_n}{r} \nu \overline{v^{2n}_{r}} + (2n-1)\nu \overline{v^{2n-2}_{r}}\frac{d\Phi}{dr}=0,
\end{equation}
which uniquely determines the radial moment if one assumes knowledge of all moments at preceding order. The definition of the separable augmented density system is then extended to $\{\beta_n\} = \beta, \;\forall n$ and thus represents the system for which the anisotropy parameters are independent of order.
We will show that calculation of the other moments then follows via a recursive passage through the remaining Jeans equations. To generalise the method used to obtain the tangential moment in equation \eqref{tanfourgen} we consider the generic Jeans equation \eqref{genjean} and introduce, where convenient, the compact notation,
\begin{equation}\label{mnot}
\nu \overline{v^{2p}_r v^{2q}_t} = m_{p,q}.
\end{equation}
and the adjacent moment ratios at order $2n=2(p+q)$,
\begin{equation}\label{momrats}
f^{p+q}_{q}(\{\beta_n\},r) \equiv (p+\frac{1}{2})\frac{m_{p,q}}{m_{p+1,q-1}} 
\end{equation}
where we note that by definition $f^{n}_{1} = 1-\beta_n$. The prefactor $p+1/2$ is included to absorb the order dependence inherent to the isotropic system \eqref{prodad} such that we can isolate the order dependence of the systems anisotropy. Rearranging equation \eqref{genjeans} for the unknown moment,
\begin{eqnarray}
(p-\frac{1}{2}) \nu \overline{v^{2p-2}_r v^{2q+2}_t} &=& \frac{r}{2}\frac{d(\nu \overline{v^{2p}_r v^{2q}_t})}{dr} + (q+1) \nu \overline{v^{2p}_r v^{2q}_t} \nonumber \\ &+& (p-\frac{1}{2})r\nu \overline{v^{2p-2}_r v^{2q}_t} \frac{d\Phi}{dr}  
\end{eqnarray}
such that upon substituting the prior moment ratio \eqref{momrats} into the derivative,
\begin{eqnarray}
\frac{dm_{p,q}}{dr} &=& \frac{1}{p+\frac{1}{2}} \left(m_{p+1,q-1} \frac{df^{n}_{q}}{dr} + f^{n}_{q}\frac{dm_{p+1,q-1}}{dr}\right) \\
										&=& \left[\frac{1}{f^{n}_{q}}\frac{df^{n}_{q}}{dr}+\frac{2}{r}(f^{n}_{q}-q)\right]m_{p,q} \nonumber -2f^{n}_{q}m_{p,q-1}\frac{d\Phi}{dr}
\end{eqnarray}
we obtain the recursive relation for $q>1$,
\begin{eqnarray}\label{recurs}
f^{n}_{q+1} &\equiv& (p-\frac{1}{2})\frac{m_{p-1,q+1}}{m_{p,q}}  \\
&=& 1 + f^{n}_{q} + \frac{r}{2}\frac{1}{f^{n}_{q}} \frac{df^{n}_{q}}{dr} \\&+& (n-q-\frac{1}{2}) \frac{m_{p-1,q}}{m_{p,q}}r\frac{d\Phi}{dr}\left[1-\frac{f^{n}_{q}}{f^{n-1}_{q}}\right].\nonumber
\end{eqnarray}
Starting with $f^{n}_{1} = 1-\beta_n$ this relation may then be used to iteratively calculate the complete set of intrinsic moments at 2$n$th order. The first iteration yields,
\begin{equation}
f^{n}_{2} = 2-\beta_n-\frac{r}{2(1-\beta_n)}\frac{d\beta_n}{dr}+ (n-\frac{3}{2}) \frac{m_{n-q-1,q}}{m_{n-q,q}}r\frac{d\Phi}{dr} \frac{\beta_n-\beta_{n-1}}{1-\beta_n}
\end{equation}
which we may use to check that with $f^{2}_{1} = 1-\beta'$ and $f^{1}_{1} = 1-\beta$, the tangential fourth moment in equation \eqref{tanfourgen} is recovered via $\overline{v^{4}_{t}} =\frac{4}{3}f^{2}_{1}f^{2}_{2}\overline{v^{4}_{r}}$. To calculate the projected moments we first note that
\begin{equation}
\frac{m_{n-k,k}}{m_{n,0}} = \prod^{q}_{j=1} \frac{m_{n-j,j}}{m_{n-j+1,j-1}} = \frac{\prod^{q}_{j=1} f^{n}_{j}}{(p+\frac{1}{2})_{q}}.
\end{equation}
and that in analogy to equation \eqref{Anal} in the separable augmented density system we may define,
\begin{equation}
\alpha^{n}_{q} \equiv \prod^{q}_{k=1} f^{n}_{k},
\end{equation}
which satisfies the recursive relation
\begin{equation}\label{algen}
\alpha^{n}_{q+1} = \alpha^{n}_{q}f^{n}_{q+1}.
\end{equation}
To check that the system converges to the separable augmented density in the limit $\{\beta_n\} = \beta$ where the order dependence is removed,
\begin{equation}\label{frec}
f^{n}_{q+1}(\{\beta_n\})\rightarrow f_{q+1}(\beta)= 1 + f_q + \frac{r}{2}\frac{1}{f_{q}}\frac{df_{q}}{dr}.
\end{equation} 
we substitute equation \eqref{frec} into equation \eqref{algen} with $\alpha_{q} = f_q \alpha_{q-1}$ to eliminate $f_q$ such that after some algebra,
\begin{eqnarray}
\alpha_{q+1} &=& \left(1+\frac{\alpha_q}{\alpha_{q-1}} +\frac{r}{2} \frac{\alpha_{q-1}}{\alpha_q} \left[\frac{1}{\alpha_{q-1}}\frac{d\alpha_q}{dr}-\frac{\alpha_q}{\alpha^{2}_{q-1}} \frac{d\alpha_{q-1}}{dr}\right]\right)\alpha_q \nonumber \\
&=& \left(1+\frac{\alpha_q}{\alpha_{q-1}} -\frac{r}{2}\frac{1}{\alpha_{q-1}} \frac{d\alpha_{q-1}}{dr}\right)\alpha_q + \frac{r}{2} \frac{d\alpha_q}{dr},
\end{eqnarray}
which one can prove inductively is satisfied by
\begin{equation}
\alpha_{q+1} = (q+k)\alpha_q + \frac{r}{2}\frac{d\alpha_q}{dr}.
\end{equation}
Applying the boundary condition $\alpha_1 = 1-\beta$ and choosing $\alpha_0=1$ yields $k= 1-\beta$ and therefore recovers equation \eqref{Anal}. To generalise the projected moments in equation \eqref{nlossep} we simply promote the ratios to $\alpha^{n}_{k}$. Whilst seemingly a subtle change, this makes the subsequent calculation considerably more cumbersome. Though the generalisation at fourth order is straightforward the higher order projected moments become rather inconvenient for practical use when one deviates from the separable augmented density system. The inclusion of the higher order analysis is however useful to ensure that the set $\{\beta_n\}$ yields positive moments at all orders and thus a physical distribution function. This constraint on the system may be represented simply as,
\begin{equation}\label{constraints}
f^{n}_{q} \geq 0,\;\;\forall \; n,q,
\end{equation}
which imposes physical constraints on the analogs including $\beta_n \leq 1$ for $q=1$ and for example the \textit{sufficient} but not strictly necessary condition for constant anisotropy $\beta_n \geq \beta_{n-1}$ corresponding to $q=2$.    
\subsection{Summary}
\begin{figure}
	\centering
		\includegraphics[width=9.5cm]{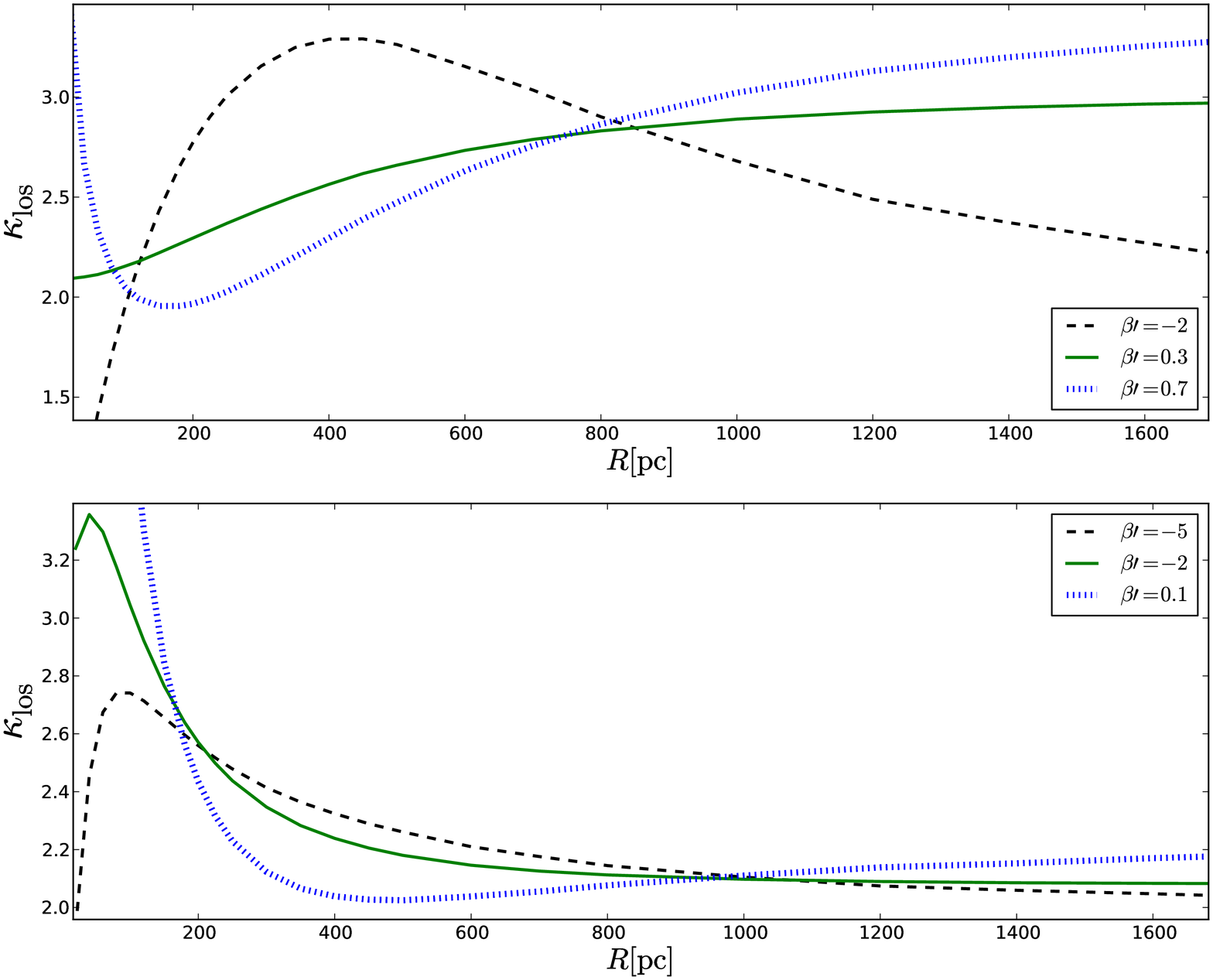}
	
	\caption{The fourth order degeneracy: Line of sight kurtosis profiles for a fixed NFW density profile and with radial $\beta=0.3$ (above) and tangential $\beta=-2$ (below) second order anisotropy. The solid green line shows Lokas's model $\beta=\beta^{\prime}$ and the blue dotted/black dashed lines show constant positive $\beta^{\prime}>\beta$ and negative $\beta^{\prime}<\beta$ deviations. For each curve we assume a Plummer stellar density profile with Plummer radius 575pc.}
	\label{fourdeg}
\end{figure}

By introducing an analog of the Binney anisotropy parameter at fourth order it is possible to represent the complete analytic set of projected fourth moments. Not only is the result presented in \cite{Lokas02} adapted in equation \eqref{LOSgenfour} to an arbitrary specification of the anisotropy parameter but as a subset of the separable augmented density system it is interpreted as the particular case where anisotropy is order-independent. To employ the method in \cite{Lokas05} to lift the Jeans degeneracy one may choose any arbitrary specification $\beta'=f(\beta)$ to construct a physical model thus removing the constraints upon the distribution function imposed by the separable augmented density. Fig. \ref{fourdeg} demonstrates that when one admits general solutions to the fourth order Jeans equations without such a constraint then there is a new degeneracy of solutions inherent to the new degree of freedom $\beta^{\prime}$. What remains to be studied is whether this degeneracy is as affecting as its traditional second order counterpart. 

 Additionally the higher order projected moments are presented such that constraints \eqref{constraints} analogous to those presented in \cite{an11b} for the separable augmented density system, $\alpha_q \geq 0$, may be used to ensure that any given model has a physical DF. Though the maths underlying the higher order moments quickly becomes impractical as one deviates from the separable augmented density system, the fourth order generalisation remains tractable. With a generic framework we facilitate a model with a stronger physical motivation and state, in general terms, a condition to lift the degeneracy namely a correlation between the anisotropy parameter analogs. Without a strong physical argument to provide this relation analytically we turn to empirical evidence. The additional freedom afforded by the anisotropy parameter analogs provides the means to test models presented in the literature. 
\section{Statistical Treatment of Discrete Velocity Data}\label{sec4}
With the extended anisotropy framework a natural progression is to devise a model for $\beta'$ with a stronger physical basis than the separable augmented density system. Due to the completely unknown nature of the anisotropy this is however a formidable task and at time of writing no such candidate has been found. As such we proceed without specification of a correlation $\beta^{\prime} = f(\beta)$ but allow both to vary freely as independent variables scanning over all solutions to the second and fourth order Jeans equations. Though as discussed in Section 3 this simply gives rise to a new degeneracy problem at fourth order, it remains to be seen whether this is as affecting as its infamous second order counterpart in a statistical analysis of velocity data.

In this section we develop an extension of the traditional Jeans analysis that jointly fits dispersion and kurtosis measurements of LOS velocity data $d$ to those predicted by the Jeans equations \eqref{LOSsecond} and \eqref{LOSgenfour}. Unlike the method proposed by \cite{Lokas05} the inclusion of $\beta^{\prime}$ makes it comparable to the second order method which scans over an unconstrained (before parameterisation) range of solutions. With suitable parameterisation of the anisotropy and density from the literature then for a set of parameters $p$ we devise a likelihood function $\mathcal{L}(d|p)$ to assess the fit. This is then used to implement an Monte-Carlo Markov Chain (MCMC) method from which one can generate posterior distributions and thus confidence intervals on $p$ and derived quantities of interest such as the mass slope and core radius.

\subsection{Anisotropy and Density Parameterisations}
To employ a likelihood function $\mathcal{L}(d|p)$ to the discrete set of line of velocity data $d$ we require a parametric form for the anisotropy and density that constitute the parameter set $p$. The Einasto density profile \citep{einasto} and its corresponding logarithmic density slope are
\begin{equation}
\rho_{\rm{dm}}(r) = \rho_{-2}\exp\left\{-\frac{2}{\alpha}\left[\left(\frac{r}{r_{-2}}\right)^{\alpha}-1\right]\right\}
\end{equation}
\begin{equation}
\gamma(r) \equiv \frac{d\ln \rho}{d\ln r} = -2\left(\frac{r}{r_{-2}}\right)^{\alpha}
\end{equation}
where $r_{-2}$ is a scale radius that indicates where the logarithmic density slope $\gamma(r)= -2$, $\rho_{-2}$ is the scale density at this radius and $\alpha$ is a constant shape parameter that determines the rate at which the density slope deviates from -2 at the scale radius. 
\begin{figure}
	\centering
		\includegraphics[width=9.5cm]{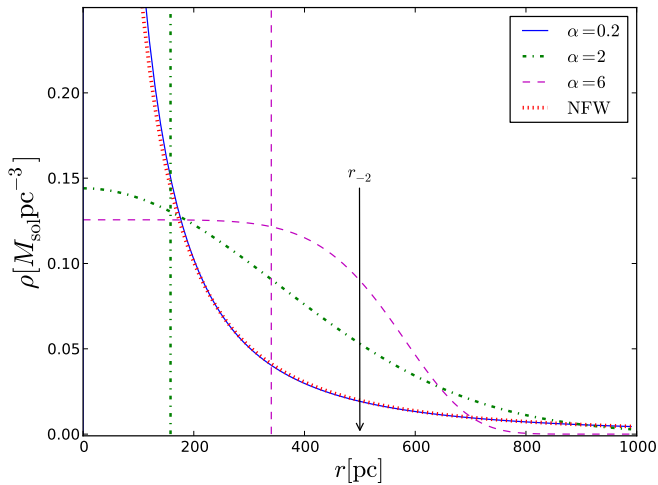}
        	\caption{Core radius of Einasto profiles: For a fixed scale radius of $r_{-2}=500pc$ (annotated with arrow) the density $\rho_{-2}$ is tuned such that the enclosed mass at 2kpc is held at a constant $M(2\rm{kpc})=10^8 M_{\rm{sol}}$ for Einasto profiles with the three shape parameters shown. The core radii $r_c$ \eqref{coreradius} corresponding to each density profile are shown as the vertical line with the same colour and line style. For $\alpha=0.2$ the vanishing core radius shrinks to 0.05pc and for display reasons is omitted from the plot. An NFW profile with scale radius of 500pc and equivalent mass at 2kpc is shown with a red dotted line for comparison. }
	\label{einasto}
\end{figure}
Though all Einasto profiles are cored as $r \rightarrow 0$, i.e have $\gamma(0)=0$, variation in the shape parameter $\alpha$ can mimic cusped profiles over the effective range of measurable distances as displayed in Fig. \ref{einasto}. If the shape parameter is small $\alpha<1$ then $\gamma$ varies slowly from the scale radius $r_{-2}$ to the centre of the galaxy, extending the steeper gradient and wiping out the core. As demonstrated in Fig. \ref{einasto} for $\alpha\approx 0.2$ the Einasto profile is very similar to the cusped ($\gamma(0)=1$) NFW \citep{NFW} profile. To get an indication of the size of the Einasto profile's core we define
\begin{equation}\label{coreradius}
\log_{10} r_{\rm{c}} = \log_{10} r_{-2} - \frac{1}{\alpha}
\end{equation}
which indicates the radius at which from zero the slope falls to  $\gamma=-0.2$ and are added to Fig. \ref{einasto} as dashed vertical lines. As expected, increasing $\alpha$ shifts the core radius towards $r_{-2}$.       

 To model the anisotropy parameters we choose not to allow radial variation in order to reduce the parameter space but with the framework established in Section 3 a model such as \citep{baes07},
\begin{equation}\label{bvh}
\beta^{(\prime)}(r) = (\beta_{\infty}-\beta_0) \frac{r^2}{r^{2}_{\beta}+r^2} + \beta_0
\end{equation}
which generically describes anisotropy with a quadratic transition about $r_{\beta}$ is also viable and with asymptotes at $\beta_0$ for $r=0$ and $\beta_{\infty}$ for $r\rightarrow \infty$ it is in line with the assumptions used to derive the kurtosis estimates \eqref{kinfty} and \eqref{kzero}. 
At second order the parameter set employed herein is thus $p_2 = \{\beta,\rho_{-2},r_{-2},\alpha\}$ which is extended at fourth order by the additional fourth order anisotropy parameter, $p_4 = \{\beta,\beta^{\prime},\rho_{-2},r_{-2},\alpha\}$. With a generalisation to radially varying anisotropy parameters this space extends considerably to $p_4 = \{\beta_0,\beta_{\infty},r_{\beta},\beta_0',\beta_{\infty}',r_{\beta}', \rho_{-2},r_{-2},\alpha\}$ which we leave for a future analysis.

Finally, a model is required for the tracer density $\nu(r)$ which, for simplicity is assumed to be a Plummer profile throughout with,
\begin{equation}
\nu(r) \propto \left(1+\frac{r^2}{r^{2}_{p}}\right)^{-\frac{5}{2}}.
\end{equation}
where $r_p$ is the Plummer radius which is equivalent to the half-light radius for this profile. This choice of $\nu$ which has a cored centre, can have a significant impact on the behaviour of the intrinsic moments at the centre of the galaxy and an analysis of real data sets should permit a greater freedom. As the primary focus here is the anisotropy, then by generating simulated data from the assumed Plummer profile we negate this effect in our analysis which is therefore optimistic in its account of uncertainties in real data.    

A calculation of the variance and kurtosis corresponding to parameter set $p$ has the following prescription. Firstly a Runge-Kutta method of order 8(5,3) (based on the `dop853' algorithm from Fortrans ODEPACK library) is used to first solve \eqref{jeans} for the intrinsic dispersion and then \eqref{radeq} for the radial fourth moment. Performing the integrals over the line of sight \eqref{LOSsecond} and \eqref{LOSgenfour} yields the LOS dispersion and kurtosis with $\kappa_{\rm{los}} = \overline{v^{4}_{\rm{los}}}/\sigma^{4}_{\rm{los}}$. 

 \subsection{Likelihood Function}
Extracting moments from the velocity data presents one of the biggest challenges of a higher order moment analysis and as highlighted in the introduction there are different ideas as to what constitutes the best approach. To utilise the Jeans equations we are limited at present to the conventional moments namely the variance and fourth moments. For realistic data sets limited sampling dominates the uncertainty associated with measurements of the moments and we discuss the relative performance of two estimators in the first subsection. To those interested predominantly in the method used to derive the results presented in Section 5 for the simulated dwarf spheroidal data sets we guide you to the concluding paragraphs of the first subsection and to the second wherein the final numerical procedure and effective likelihood function is described. 

\subsubsection{Choice of Estimator}

 For a data set $d$ of $N_{\star}$ velocity measurements $\{v_i\}$  with associated experimental errors $\{\delta_i\}$ then to maximise information one would like to choose the likelihood function to be simply the product of individual probability density functions for each tracer. The issue of this approach is that it is difficult to ensure that, without \textit{explicitly} incorporating errors from limited sampling, the uncertainties are reliable. As such we split the data into radial bins for which we fix $\sigma_p, \kappa_p$ at one radius and use estimators of the variance and kurtosis of each bin, $\widehat{\sigma}^2$ and $\widehat{\kappa}$ whose sampling distribution will of course depend on the size of the bin chosen. In this way the effect of limited sampling is manifested as loss of statistical precision in small bins and a loss of spatial resolution for large bins. The likelihood function is then,
\begin{equation}\label{truelike}
\mathcal{L}(d|p) = \prod^{N_r}_{i}\mathcal{S}(\widehat{\sigma}^{2}_{i},\widehat{\kappa}|\sigma^{2}_{p}(R_i),\kappa_p(R_i))
\end{equation}
where $N_r$ is the number of radial bins, $R_i$ is the mean radius of the $i^{th}$ bin and $\mathcal{S}$ is the joint sampling distribution of estimators dependent upon the sample size $N$ and the proposed parameters $p$. The form of this distribution is calculated numerically by drawing many bootstrap samples of $N$ velocities from a parent distribution $\mathcal{F}(v_i|\sigma^2_p,\kappa_p)$ such as those outlined in Appendix B for which the sample moments converge to $\sigma_p$ and $\kappa_p$ for very large $N$.

Under the assumption that the sampling distributions are approximately normal and that the estimators are independent then the likelihood \eqref{truelike} may be modelled as the product of two $\chi^2$ distributions with each estimator measurement distributed normally and centred, in the absence of bias, at $(\sigma_p,\kappa_p)$. In this ideal scenario then there is a simple analytic form for the likelihood,
\begin{equation}\label{likechi}
\mathcal{L}_\chi(d|p) = \prod^{N_r}_{i} \mathcal{S}^{\chi}_{\sigma^2}(\widehat{\sigma}^{2}_{i}|\sigma^{2}_{p}(R_i),\kappa_p(R_i)) \times \mathcal{S}^{\chi}_{\kappa}(\widehat{\kappa}_i|\sigma^{2}_{p}(R_i),\kappa_p(R_i))
\end{equation}
where for moment estimator $\widehat{m} = \{\widehat{\sigma}^2, \widehat{\kappa}\}$,
 \begin{equation}
\mathcal{S}^{\chi}_{m}(\widehat{m}|\sigma^{2}_{p},\kappa_p) = \frac{1}{\sqrt{2 \pi \rm{Var}(\mathcal{S}^{\chi}_{m})}} \exp\left( -\frac{(\widehat{m} - m_p)^2}{2 \rm{Var}(\mathcal{S}^{\chi}_{m})}\right).
\end{equation}
As such to assess the merit of an estimator we consider not only the convergence to the true value that minimises $\rm{Var}(\mathcal{S}^{\chi}_{m})$ for optimal statistical precision but also the ability to effectively model the bias $b=\widehat{m}-m_p$ and deviations from Gaussianity. Further consideration must also be given to how robust the sampling distributions are to the choice of parent distribution $\mathcal{F}$ from which the bootstrap samples are generated.

We considered two estimators, namely the standard sample moments,
\begin{equation}\label{sampdisp}
\widehat{\sigma}^{2}_{S} = \frac{1}{N} \sum^{N}_{i=1} (v_i - \mu )^2
\end{equation}
\begin{equation}\label{sampkurt}
\widehat{\kappa}_S = \frac{1}{N} \sum^{N}_{i=1} \frac{(v_i - \mu )^4}{(\widehat{\sigma}^{2})^2}
\end{equation}
where $\mu$ is the mean velocity that is set to zero for a rotationless system and maximum likelihood estimators $\widehat{\sigma}^{2}_{M}$ and $\widehat{\kappa}_{M}$ defined by ,
\begin{equation}
 \prod^{N}_{i} \mathcal{F}(v_i|\widehat{\sigma}^{2}_{M},\widehat{\kappa}_{M}) = \rm{max} \left\{ \prod^{N}_{i} \mathcal{F}(v_i|\sigma^2,\kappa)\right\}
\end{equation}
where the family of distributions $\mathcal{F}(v|\sigma^2,\kappa)$, not necessarily equivalent to the true parent distribution from which samples are generated $\mathcal{F}_{\rm{true}}$, is a prior estimate. A Nelder-Mead algorithm is used to find the parameters in $\mathcal{F}_M$ that maximise the bins likelihood. 

For both estimators we generated, from both families in Appendix B, many samples of various sizes encompassing a wide range of input variance and kurtosis parameters.  Experimental errors, assumed to be normally distributed with variance $\delta^2$, were added to the velocities.  To account for experimental errors in the maximum likelihood function a convolution \citep[see][for details]{Amorisco}  with the Gaussian distribution of errors  $\mathcal{F} \to \mathcal{F} * \mathcal{G}(0,\delta)$ was used to account directly for the noise. Calculating the estimators for each sample the sampling distributions were constructed. For samples with $N<400$ velocity measurements, which is the maximum that one can realistically expect in a radial bin for current dwarf spheroidal data sets, the maximum likelihood estimator far outperformed the sample kurtosis in reducing bias $\kappa - \widehat{\kappa}$ and achieved a better precision about the parent value $\kappa$. A major problem however is that the marginalised distribution of both kurtosis estimators $\mathcal{S}_\kappa$ is increasingly skewed as one raises $\kappa$ such that it isn't valid to assume a chi-squared form \eqref{likechi} for the likelihood function. To ameliorate this effect we transformed the kurtosis estimator as prescribed in \cite{Lokas05} to $\widehat{\kappa^{\prime}} = (\ln\widehat{\kappa})^{0.1}$ but despite a significant improvement the sampling distributions from $\kappa_p$ are not adequately approximated by a normal distribution with the parent distributions considered and $N<400$. We were also unable to establish a sufficiently accurate fit with a skew-normal distribution. 

Following the tests we thus came to the following conclusions. Firstly, to model the sampling distribution it is not always appropriate to assume that the estimators are normally distributed with samples generated from leptokurtic distributions demonstrating significant skew for all estimators. As such we elect not a likelihood function based on the chi squared distribution \eqref{likechi} (for which one needs only the bootstrap variances to approximate the sampling distribution) but to construct the sampling distribution $\mathcal{S}$ purely numerically.

Secondly, as it can take as many as $10^6$ bootstrap samples to achieve a smooth distribution after binning, probing the two parameter phase space for maximum likelihood estimators $\widehat{\sigma}^{2}_{M}$ and $\widehat{\kappa}_M$ is too computationally expensive for practical use. For this reason, in spite of the inferior statistical precision, we choose the sample moments $\widehat{m}_S$ as our estimator. Building sample distribution look-up tables for a large range of kurtosis values with $10^6$ maximum likelihood evaluations just takes too long and once committed to a fully numerical probability distribution the significant bias of $\widehat{\kappa}_S$ is also accounted for.

\begin{figure*}
	\centering
		\includegraphics[width=17cm]{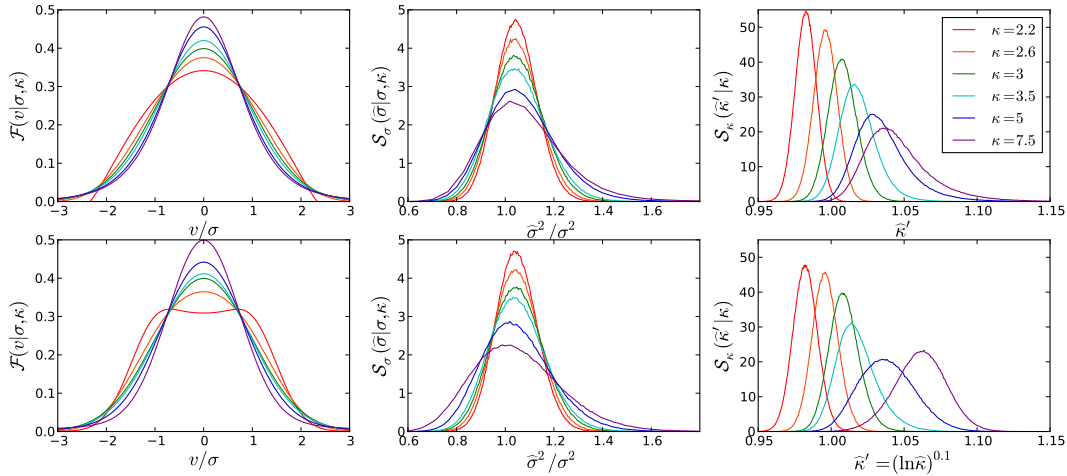}
        	\caption{Sampling distributions for Pearson (top panel) and Gaussian Superposition (bottom panel) PDFs: On the left are the parent distribution functions with fixed unit variance and kurtosis as displayed with the green curve a perfect Gaussian and shift towards red and blue being increasingly platy/leptokurtic. Middle and right columns are the sampling distributions (after $10^6$ tests) of the estimators \eqref{sampdisp} and \eqref{sampkurt} as evaluated for a sample of 200 stars drawn from the distributions in the left column with experimental errors drawn from a Gaussian with width $\delta = 0.22 \sigma$ which accounts for the bias in sample variance. To aid readers of a black and white version of this figure we note that increasingly leptokurtic distributions exhibit systematically larger central values for the parent distribution $\mathcal{F}$, wider sampling distributions of variances $\mathcal{S}_{\sigma}$ and sampling distributions of the kurtosis estimator $\mathcal{S}_{\kappa^{\prime}}$ that are increasingly shifted to the right. A colour version of this figure is available online.  }
	\label{sampdists}
\end{figure*}

In Fig. \ref{sampdists} the marginalised sampling distributions are shown for both parent distribution families used in our assessment for which formal definitions are given in Appendix B. For clarity we only display the distributions for the sample moments $\widehat{m}_S$ where sufficient tests can be made in a limited time to generate smooth curves. Inspection of the curves shows the increasing skewness of both the variance and kurtosis estimator distributions as one samples from the leptokurtic distributions with longer thinner tails. It is these distributions that also show greatest variation to the choices of parent distribution considered. By contrast though the most platykurtic choice $\kappa = 2.2$ displayed in red for the Pearson and Gaussian superposition families have rather distinct parent distributions the resultant sampling distribution is rather similar in shape. The finite support of the Pearson distribution is manifested by a slightly narrower sampling distribution $\mathcal{S}_{\kappa^{\prime}}$ than that of the Gaussian superposition that permits all velocities. In this sense the rather common feature of platykurtic distributions to have finite support, which is troublesome for maximum likelihood estimators \footnote{Though naturally catered for by scaling the variance with the kurtosis as in the Henon-Lokas model shown in Fig. \ref{lokasdists}}, has much less impact here.

\subsubsection{Numerical Evaluation of the Likelihood}

Having chosen to use radial bins and the sample moment estimators \eqref{sampdisp} and \eqref{sampkurt} to fit the data, we now outline the process by which we numerically evaluate the joint sampling distribution $\mathcal{S}(\widehat{\sigma}^2,\widehat{\kappa}|\sigma^2_p,\kappa_p)$ that for a proposed set of parameters $p$ yields the probability for a given pair of variance and kurtosis estimator measurements that feeds into the likelihood \eqref{truelike}.

For simplicity we assume that the estimators are independent such that the likelihood function may be separated but in practice both the sample moment estimators are weakly positively correlated with a Pearson coefficient of $\rho \approx 0.2$. This is also true of the maximum likelihood estimators that we considered. To a much better approximation, the marginalised distribution of kurtosis estimators is independent of the variance of the parent distribution $\sigma^{2}$ and the marginalised distribution of variance estimators $\mathcal{S}_\sigma(\widehat{\sigma}^2|\sigma^2,\kappa)$ is also approximated excellently by
\begin{equation}\label{kdrop}
\mathcal{S}_\sigma(\widehat{\sigma}^2|\sigma^2_1,\kappa) \approx \left(\frac{\sigma_2}{\sigma_1}\right)^2 \mathcal{S}_\sigma\left(\frac{\sigma^2_1\widehat{\sigma}^2}{\sigma^2_2}|\sigma^2_2,\kappa\right)
\end{equation}  
which is exactly true if $\mathcal{S}_\sigma$ is a perfect Gaussian. By choosing an appropriate scale for the variance then the sample distributions can effectively be built by varying only the kurtosis. With these assumptions and changing to the transformed sample moment kurtosis estimator $\widehat{\kappa}^{\prime}$ the effective likelihood function \eqref{truelike} becomes
\begin{equation}\label{like}
\mathcal{L}(d|p) = \prod^{N_r}_{i} \mathcal{S}_\sigma(\widehat{\sigma}^2_i|\sigma^2_p(R_i),\kappa^{\prime}_p(R_i)) \times \mathcal{S}_\kappa(\widehat{\kappa}_i^{\prime}|\kappa^{\prime}_{p}(R_i)).
\end{equation}
In practice then for a data set of $N_\star$ velocity measurements we split the data into $N_r$ radial bins of equal tracer (e.g stellar) content such that there are $N = N_\star / N_r$ tracers in each bin. For each bin we calculate the estimators of variance $\widehat{\sigma}^2$ and kurtosis $\widehat{\kappa}^{\prime}$. Selecting a variance scale $\sigma^{2}_{s}$ of a similar magnitude to the variance estimators we draw many velocity samples of $N$ stars from parent distributions $\mathcal{F}(v|\sigma^2_s,\kappa)$ for a range of kurtosis values that we pessimistically expect to encompass the true values. For each of these bootstrap samples we add normally distributed noise to mimic experimental error and then calculate the estimators \eqref{sampdisp} and \eqref{sampkurt} such that with many ($\approx 10^6$) samples we can construct the sampling distributions $S_\sigma$ and $S_\kappa$ corresponding to the relevant parent distribution. The sampling distributions, which provide the probability of observing a given measurement of the estimators, are then stored as look-up tables for different discrete values of kurtosis. Upon selecting a parameter set $p$ the two relevant sampling distributions encompassing $\kappa_p$ are found from the look-up tables and a linear interpolation of the relevant probabilities returns the likelihood of the estimator measurements.

\subsection{MCMC Methodology and Output Analysis}
With the likelihood function \eqref{like}, the marginalised posterior distributions for each parameter in $p$ are efficiently generated in higher dimensional parameter spaces by creating a Monte-Carlo Markov Chain (MCMC) wherein the normalised density of entries for parameter $p_i \in p$ in a region $a<p_i<b$ is, assuming that the chain has converged, the probability that $p_i$ lies in that region. To construct the MCMC the Metropolis-Hastings \citep{metro} algorithm is used to iteratively generate new parameter space locations $p^{\prime}$ from the current location $p$ from a proposal density $\mathcal{Q}(p^{\prime}|p)$ and then accept or reject the inclusion of $p^{\prime}$ according to the ratio of likelihoods $r=\mathcal{L}(d|p^{\prime})/\mathcal{L}(d|p)$. If $r>1$ then as a more likely set of parameters $p^{\prime}$ is always accepted to the chain and the location is updated to $p=p^{\prime}$ whilst if $r<1$ then $p^{\prime}$ is accepted if and only if $r>U$ where $U\in[0,1]$ is generated randomly from the uniform distribution. When this latter condition is not met then $p^{\prime}$ is not added to the chain and another parameter set is drawn from the proposal density centred at $p$. The acceptance ratio must be balanced to ensure both sufficient exploration of the parameter space and ability to precisely determine the best fit regions. In our analysis we adopt ensure that all chains have an acceptance ratio of between 0.2 and 0.3 in line with that recommended 
for higher dimensional models . 

With this random walk then adjacent iterations of the chain are highly correlated and one often requires many iterations before a convergence of the posterior distributions is achieved. The rate of convergence is optimally efficient if the proposal density is close to that which is being measured. For this reason the data is used to shape the proposal density and we refer the reader to the comprehensive guide in the appendices of \cite{Lewis}. Firstly we vary one parameter at a time with a Gaussian one dimensional proposal density centred at the previous chain location and with fixed width $T_i$ referred to as a temperature. The temperature is then adjusted until the chain has an acceptance of roughly $50\%$. We then enter these $N_p$ temperatures as the diagonal elements of an otherwise zero $N_p \times N_p$ covariance matrix that forms the intial estimate an $N_p$ dimensional multivariate normal proposal density centred at $p$. An additional global scale is introduced to the covariance matrix to ensure that the acceptance ratio can be set in the desired range. After a set number of iterations we then periodically update the covariance matrix to learn the covariances and note a significant improvement in the mixing due to the significant correlation between for example the density parameters. With a multivariate Gaussian proposal density we also found it beneficial to transform the parameters with appropriate log scaling and in line with \cite{charbonnier2011} we adopt the following flat priors and ranges,
\begin{eqnarray}\label{priors}
\log_{10}[1-\beta]&:&[-1,1] \nonumber\\
\log_{10}[1-\beta^{\prime}]&:&[-1,1] \nonumber\\
\log_{10}\rho_{-2}&:&[-8,3]\\
\log_{10}r_{\text{-2}}&:&[1,5]\nonumber \\
\log_{10}\alpha&:&[-1.3,1.2]\nonumber.
\end{eqnarray}

 For the final chain the covariance matrix is fixed to ensure that the chain is Markovian and we remove a significant percentage of the initial points as a burn in period wherein the chain moves to the high likelihood region.

To assess the mixing and convergence of the output chains we perform a number of visual and statistical tests. Trace plots of the walks indicate the quality of mixing and we verify this by calculating the auto-correlation function. The convergence length of the chain is defined as the maximal lag amongst all parameter chains at which the auto-correlation function dips below 0.5. Typical convergence lengths are about 25 though for some they extend to 50 and we use this number to thin the chains before analysis such that each entry is approximately independent. As a minimal requirement we set one condition that the total length of the chain is at least two thousand times the correlation length with total chain lengths of order $10^5$ iterations. Plots of the running median and quantiles are used to visually assess the convergence. With simulated data, for which we have the benefit of knowing the true underlying values, we did not (due to resource and time restraints) run  multiple chains for every data set which is the best indicator of convergence. Such a test was carried out with parallel chains with different starting locations for data sets B and G (see section 5) and we found that each parameter had a Gelman-Rubin test value \citep{Gelman92} in the final chains of $R<1.03$. For a real data set this would be a necessary practice when one has no guarantee that the chains have found the regions of highest likelihood. 

From the chains we are able to construct probability distributions and thus confidence intervals for a number of interesting physical quantities such as the core radius \eqref{coreradius} and the logarithmic mass slope $\Gamma$. Calculation of these quantities are calculated for each value of iteration of the chain. The median values of these new chains then correspond to best estimates and one may use fractiles to determine the central confidence intervals, i.e if $ 95\% $ of $\Gamma$ values in the chain lie between $\Gamma_u$ and $\Gamma_d$ then these become the $ 95\% $ central confidence intervals.  

\section{Application to Simulated Dwarf Spheroidal Galaxies}

\subsection{Simulated dSph Input Parameters}
Recently kinematic surveys \citep{walkdat} of the classical dSphs have expanded the number of velocity measurements to the order of $10^3$ which reduces the fourth order statistical errors to more tolerable levels. To test the statistical method described in Section 4 we generate simulated data sets with known parameters and see how well they are recovered. For comparison the same data sets are analysed without reference to the kurtosis. In this second order analysis the full likelihood function \eqref{like} is reduced to,
\begin{equation}\label{likesec}  
\mathcal{L}_\sigma(d|p) = \prod^{N_r}_{i} \mathcal{S}_\sigma(\widehat{\sigma}^2_i|\sigma^2_p(R_i),\kappa=3).
\end{equation}
With a key motivation of the analysis being to distinguish between shallow profiles with extended cores and those with steeper inner slopes that within the Einasto family are the best approximation to a cusped NFW-type profile, we choose two different sets of parameters (Table. \ref{simparams}) to exhibit these properties. Each has a DM density described by the Einasto profile and constant anisotropy parameters $\beta$ and $\beta^{\prime}$ which crucially are allowed to freely vary thus providing the most general description of anisotropy from a Jeans analysis to date. Additionally we sought to choose parameters that yielded realistically flattish dispersion profiles that are difficult to distinguish with the variance data alone. By choosing parameter sets with distinct anisotropy, which for the parameterisations chosen has a particularly  pronounced effect at the centre of the galaxy, we optimistically gave the fourth order method more chance to distinguish the two.  
\begin{table}
\caption{Simulated Data Parameters}
\centering
\begin{tabular}{c c c c c c}
\hline
Data set & $\beta$ & $\beta^{\prime}$ & $\rho_{-2}\;[M_{\rm{sol}}\rm{pc}^{-3}]$ & $r_{-2}\; [\rm{kpc}]$ & $\alpha$  \\
\hline
Shallow A-D & 0.25 & 0.2 & 0.05 & 1 & 3.5  \\
Steep E-H & -0.7 & -0.6 & 0.06 & 0.6 & 0.3  \\
\hline
\end{tabular}
\label{simparams}
\end{table}
As limited sampling is the dominant source of uncertainty four independent data sets are drawn from each set to explore this effect with A-D from the shallow and E-H from the steep parameter sets. Each data set comprises $2000$ tracers in line with the largest data sets (Fornax $\approx 2400$, Sculptor $ \approx 1400$) currently available for dwarf spheroidal LOS stellar velocity data. For all data sets a Plummer profile with half-light radius $r_p =575$pc is assumed and used to generate the LOS radial coordinates for each star $R_i$. To generate the velocities we calculate the line of sight variance and kurtosis profiles corresponding to the relevant parameters in Table. \ref{simparams} which for each star will yield the relevant values $\sigma^{2}_{p}(R_i)$ and $\kappa_p(R_i)$. From these we then draw each stars velocity $v_i$ from the distribution $\mathcal{F}(v|\sigma^2_p(R_i),\kappa_p(R_i))$ by selecting a uniform random number and inputting it into the inverse cumulative density function of $\mathcal{F}$. To simulate the possible discrepancy that may arise from estimating $\mathcal{F}$ incorrectly, we draw samples for the data sets from the Gaussian superposition functions in (Appendix B and Fig.\ref{sampdists}) whilst assuming the Pearson form for the sampling distributions $\mathcal{S}$ used in the likelihoods \eqref{like} and \eqref{likesec}. Experimental error velocities distributed by a Gaussian of width $\delta = 0.22 \sigma$, in line with Fornax \citep{Amorisco} are then added to the samples for the final data set.      
\begin{figure}
	\centering
		\includegraphics[width=9cm]{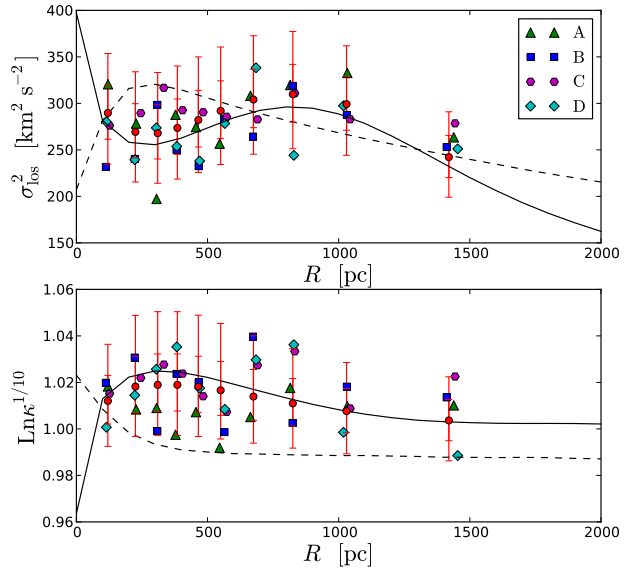}
	
	\caption{Simulated galaxies with extended cores: Moment estimators \eqref{sampdisp} and \eqref{sampkurt} are plotted for galaxies A-D for 10 radial bins of 200 stars. The black curves show the line of sight dispersion $\sigma^{2}_{p}(R)$ and kurtosis $\kappa_p(R)$ as calculated from \eqref{LOSsecond} and \eqref{LOSfour} with the solid line corresponding to the shallow parameters (see Table.\ref{simparams}) from which the data A-D are generated and the dashed line showing for reference the steep parameter set (from which E-H are generated). The error bars show the median (central circular data point), $68\%$ and $95\%$ confidence intervals derived from the sampling distributions $\mathcal{S}$ pertaining to the shallow parameter set and solid black curve.} 
	\label{coregals} 
\end{figure}

\begin{figure}
	\centering
		\includegraphics[width=9cm]{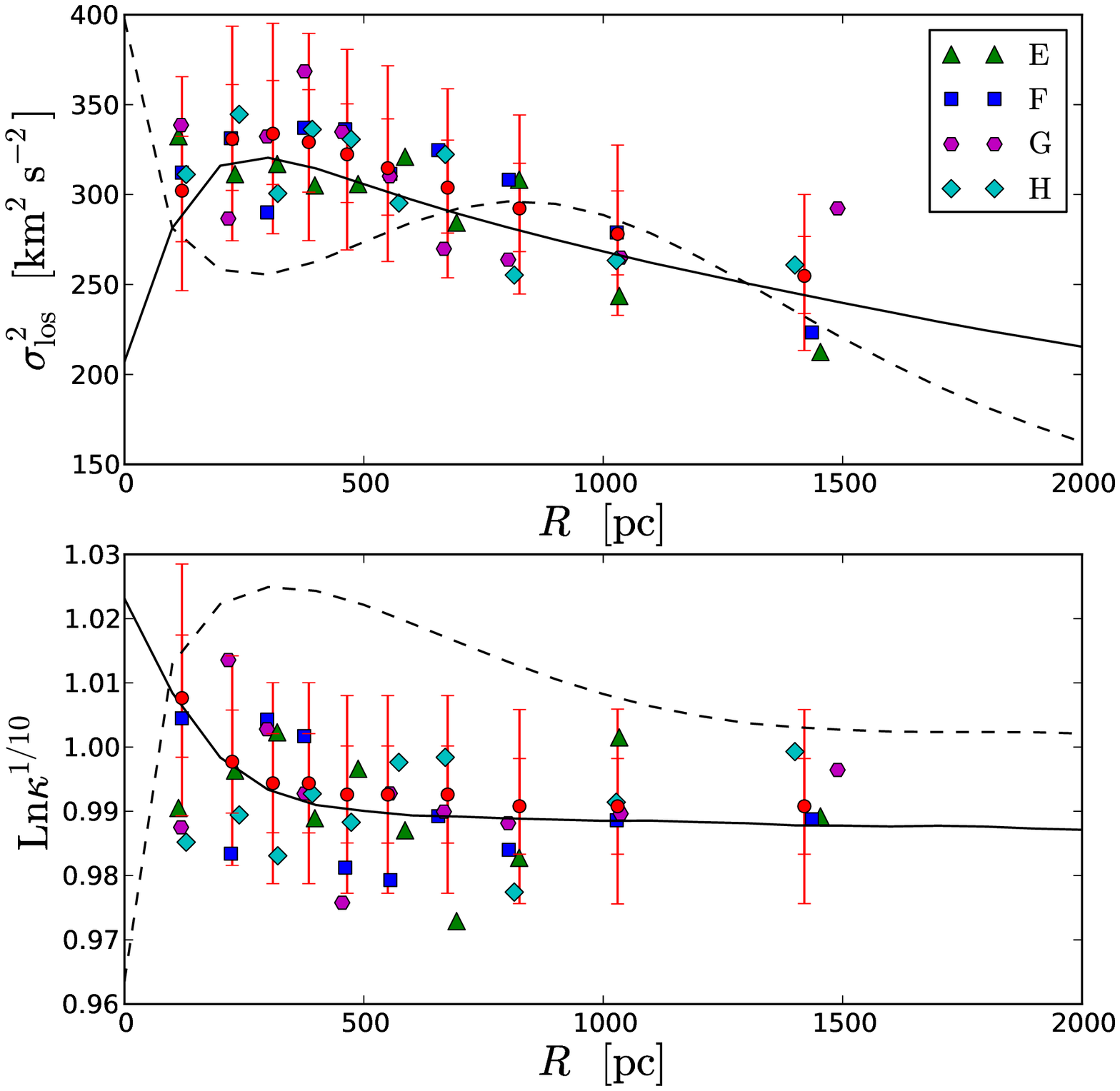}
	
	\caption{Simulated galaxies with steep density profiles: Key as per Fig.\ref{coregals} but with galaxies E-H generated by the steep parameter set with corresponding errors. Again the solid black line shows those parameters from which the data is generated and the dashed line shows the alternative for reference.} 
	\label{cuspgals} 
\end{figure}

Figs. \ref{coregals} and \ref{cuspgals} show the sample moments of the binned simulated data sets and the considerable scatter from limited sampling and experimental error about the curve corresponding to the input parameters. The errors are representative of the sampling distributions which are generated solely from the parameters $p$ and no reference to the data. The experimental errors systematically increase the predicted dispersion and smear out the kurtosis to Gaussianity with decreases to the leptokurtic shallow parameter curve and increases to the platykurtic steep counterpart. The smaller uncertainty in sample kurtosis from platykurtic parent distributions (increasingly red in Fig. \ref{sampdists}) results in a smaller scatter amongst the steep galaxies E-H which also benefit from a flatter kurtosis profile that minimises the radial information lost from only having ten radial bins. The distinct choice of anisotropy parameters, to which the kurtosis is particularly sensitive, appears visually to offer a distinction between the two parameter sets in spite of the large sampling uncertainties.         

\subsection{MCMC Results}

For each simulated dwarf galaxy the MCMC analysis was performed as described in Section 4  for parameters $p=\{\beta,\beta^{\prime},\rho_{-2},r_{-2},\alpha\}$ varying freely in the ranges \eqref{priors}. 

	

\begin{figure*}
	\centering
		\includegraphics[width=17cm]{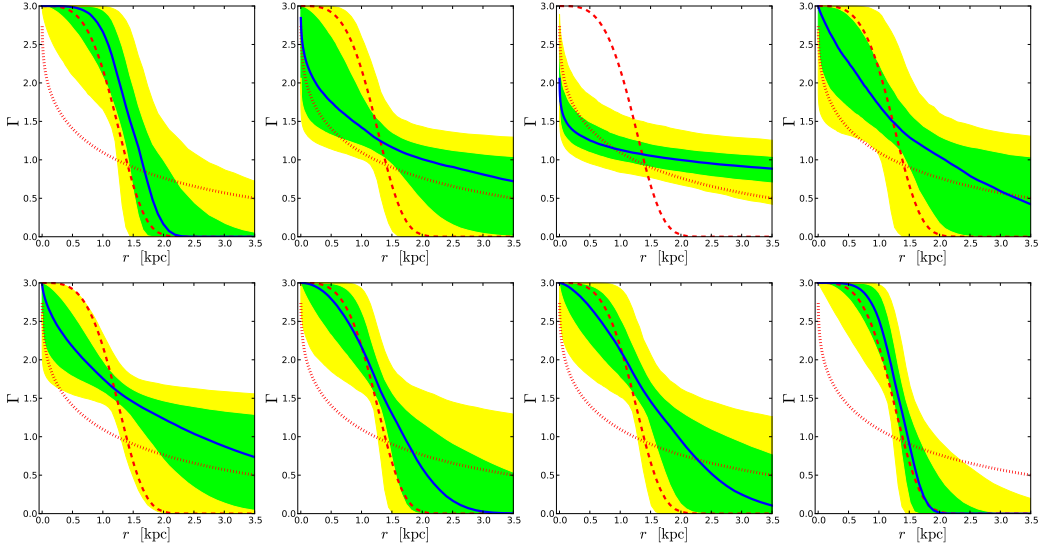}
	
	\caption{Mass slopes of shallow (extended core) galaxies A-D: The top panel shows the median (solid blue line) and $68\%$ (green) and $95\%$ (yellow) confidence regions returned by the MCMC output for the dispersion only analysis with \eqref{likesec} whilst the bottom panel shows the MCMC output with the additional kurtosis information and full likelihood \eqref{like}. The dashed red line indicates the mass slope derived from the shallow input parameters (Table. \ref{simparams}) from which datasets A-D are generated and for reference in dotted red is the steeper parameters from which data sets E-H are generated.} 
	\label{coreg} 
\end{figure*}
\begin{figure*}
	\centering
		\includegraphics[width=19cm]{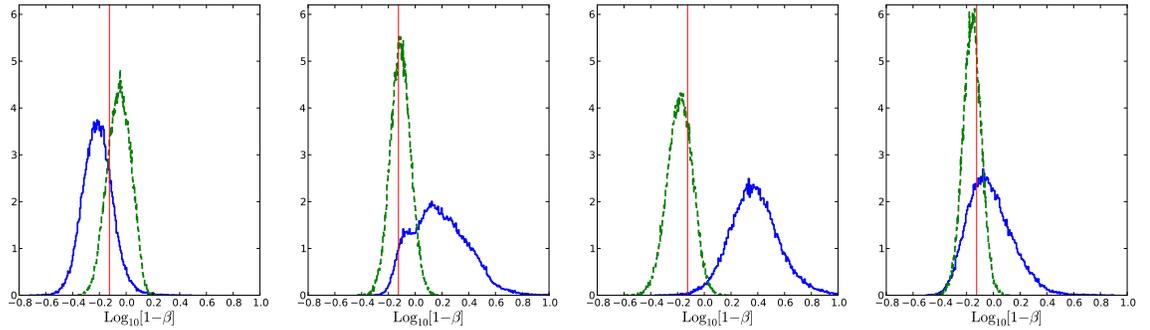}
	
	\caption{Second order anisotropy $\beta$ of data sets with shallow input parameters A-D: Posterior distributions of the anisotropy parameter are shown for the dispersion only MCMC analysis (blue solid) and dispersion-kurtosis joint analysis (green dashed). The true value of $\beta$ from which the data are generated is indicated by a solid red line.} 
	\label{corea} 
\end{figure*}
\begin{figure*}
	\centering
		\includegraphics[width=17cm]{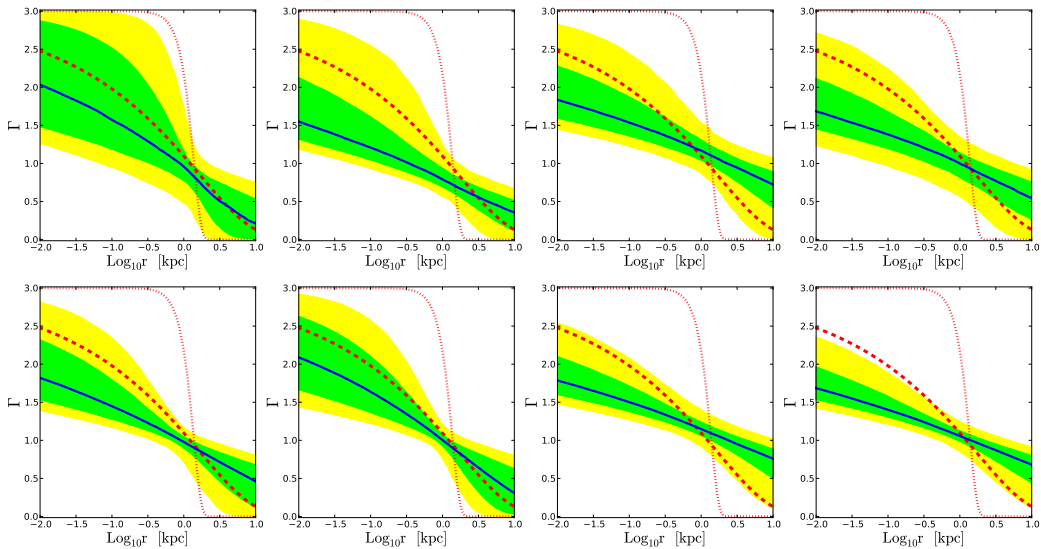}
	
	\caption{Mass slopes of steep (cusp-like) galaxies E-H: Key as per Fig. \ref{coreg} but to highlight the behaviour at the centre of the galaxy a log scale is used for the horizontal axis. In this instance the dashed line corresponds to the steep parameters with the dotted line showing the shallow parameters for comparison.} 
	\label{cuspg} 
\end{figure*}
\begin{figure*}
	\centering
		\includegraphics[width=19cm]{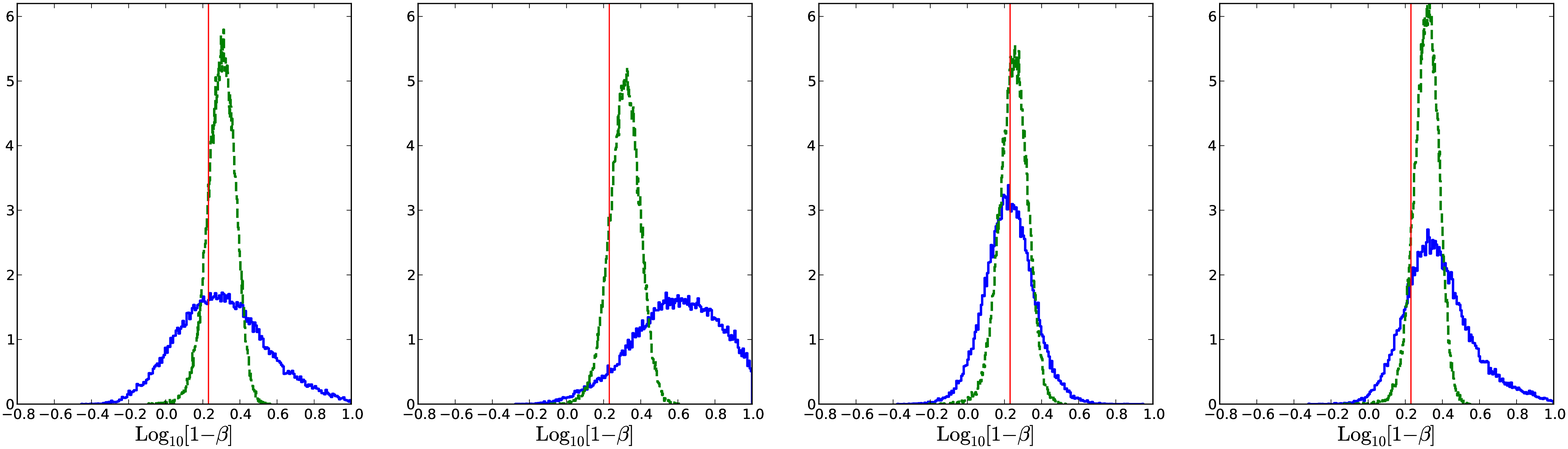}
	
	\caption{Second order anisotropy of galaxies E-H: Key as per Fig. \ref{corea}.} 
	\label{cuspa} 
\end{figure*}
Rather than plotting the posterior distributions for the individual density parameters we calculate the mass slope $\Gamma = d \ln M/d \ln r$ for each element of the chain and over a range of radii. These $\Gamma$ chains are then sorted and the quantiles used to determine the confidence intervals at each radius. Mass slope plots, which indicate whether a galaxy exhibits core-like $\Gamma(R\to0) = 3$ or cusp-like $\Gamma(R\to0) = 2$ behaviour at the centre, enable a direct comparison with measurements obtained with analysis of multiple populations \citep{penarrubia} that place strong constraints in this regard. This aspect of the degeneracy problem is at the heart of the motivation for this paper with implications for the expected flux of DM in indirect detection and tensions with $\Lambda$CDM simulations. We also highlight the impact of incorporating the kurtosis on the anisotropy parameter $\beta$ which is poorly constrained in a dispersion-only analysis.

The first reassuring result is that the MCMC analysis recovers the true parameters within the stated $95\%$ confidence intervals every time with the notable exception of the second order treatment of galaxy C for which the galaxies extended core is completely excluded. On studying Fig. \ref{coregals} we see that the cluster of variance measurements below four hundred parsecs all lie above the median value and thus by chance favour a downturned dispersion curve with tangentially biased anisotropy which is clear in Fig. \ref{corea}. In the fourth order analysis however the leptokurtic sample points come to the rescue and in spite of the anomalous variance measurements the MCMC faithfully recovers the input value. This highlights the role of the kurtosis as a consistency check as means to improve the accuracy as well as the precision of the analysis. As a caveat however we note that this analysis with only 10 variance measurements is not competitive with the most recent dispersion-only analyses in the literature for which the use of 30 or 40 radial bins to measure the variance is a statistically viable way to eliminate these resolution effects. In a converse situation galaxy A has a cluster of unexpectedly low measurements of kurtosis which explains why the fourth order fit is not only worse than its second-order counterpart but less precise. In this instance though the joint analysis MCMC still returns the true value (albeit just about) within the $95\%$ confidence bands. For the more uniformly spread B and D galaxies we see that the considerable improvement in precision of the anisotropy is accompanied by an improvement in precision of Gamma wherein the fourth order breaks the degeneracy between the two parameter sets favouring the extended core to a high significance.

A study of the steep profile shows a considerable enhancement in precision. In this instance however we see that this is echoed in the second order analysis and thus suggests that the parameters chosen either do not yield a sufficiently flat dispersion profile to recreate the degeneracy inherent to real dSphs or that generating and fitting the data from a limited range of einasto profiles and constant anisotropy is not realistically varied. The fourth order analysis again improves the precision which breaks the degeneracy for galaxy E. 
In the other galaxies however the precision seems to be overestimated and the median value regularly underestimates the true value in Fig. \ref{cuspg}. This less reliable estimate could be a result of the long tails in the posterior distributions for steep density profiles which we confirm from the analysis in \cite{charbonnier2011} for which higher quantiles takes longer to converge and which are less well described with a Gaussian proposal density. 

In conclusion we find that the fourth order method returns the true anisotropy and density parameters within $95\%$ confidence in all but one instance (galaxy H) which falls just outside at small radii. This is in contradiction to the second order analysis that, though perhaps weakened relative to the best analyses in the literature due to fewer radial bins, is more susceptible to spurious statistical fluctuations from limited sampling. In three instances, with arguably the largest and  most uniform scatter, the fourth order method breaks the degeneracy between profiles with extended and very short length cores that mimic NFW profiles at all but the smallest radii. In analyses of real data no dispersion-only Jeans analysis has demonstrated a significant exclusion of vanishingly small or extended cores and thus the simulated data sets (B,D and E) wherein this is true are arguably the most realistic. It is these simulated galaxies that gain most from the inclusion of the kurtosis.

\section{Discussion}

The Jeans analysis is extremely useful in identifying the gravitational potential from the line of sight velocities of stars moving in that potential and is used to learn more about many different kinds of astrophysical objects.  The information obtained in this way is limited by degeneracies which exist as a result of our ignorance of the velocity anisotropy within the stellar tracer populations. Following \cite{Lokas02} we looked not only at the width (2nd moment of velocity) of the stellar velocities but also their kurtosis (ratio between 2nd and 4th moment) in a bid to break some of these degeneracies. With increasingly large samples we are also motivated to consider the higher order moments because the assumption of Gaussianity is inconsistent with an equilibrium solution. 

 A major problem with utilising the two fourth order Jeans equations however is that without additional information about the fourth moments then the system is under constrained, i.e there is no unique prediction for the fourth moment of the LOS velocity distribution given the anisotropy parameter $\beta(r)$ and density $\rho(r)$ alone. In the literature this has been addressed by assuming a particular form for the distribution function such as the separable augmented density wherein the ratio of fourth order moments is correlated with $\beta$ the ratio of second moments. The nature of this assumption is, from a physical standpoint, arbitrary and restricts the range of solutions such that systems with tangentially biased variances ($\beta<0$) must necessarily have tangential velocity distributions with flatter tops than their radial counterparts.

In this paper we have presented a new mathematical framework for calculating the higher order moments of the Jeans equation based upon introducing an analogue of the Binney Anisotropy parameter at each higher order and we have demonstrated that this determines the complete set of solutions to the Jeans equation at each order. At fourth order it is shown that the introduction of the analog $\beta^{\prime}$ allows for a free variation of the shape parameters with $\beta^{\prime}>0$ naively implying a more flat topped tangential distribution and $\beta^{\prime}<0$ implying a more flat topped radial distribution. With the fourth order anisotropy $\beta^{\prime}$ as an independent parameter one can scan the entire range of distribution functions in a likelihood analysis and we show that not only is the necessary extension to the Jeans equations and projected moments straightforward but by design the separable augmented density system is the limiting case $\beta^{\prime} = \beta$.

 With no \textit{a priori} intuition for a correlation between the anisotropy parameters $\beta^{\prime} = f(\beta)$ however the degeneracy problem is not guaranteed to be improved as the additional free parameter simply introduces a new degeneracy in the kurtosis measurement. To test whether this new degeneracy was as affecting as its notorious second order counterpart we developed a method to compare the constraints on density parameters for eight simulated dwarf spheroidal data sets with sample size, experimental errors and stellar surface density comparable to Fornax the largest existing data set. Einasto profiles were assumed for the density of dark matter with four A-D exhibiting an extended inner core and the other four E-H having a vanishing core that mimics an NFW at resolvable distances from the galactic centre. Under the assumption of constant anisotropy $\beta$ and for the first time introducing an independent constant $\beta^{\prime}$ to minimally close the fourth order Jeans equation we performed a traditional dispersion-only and dispersion-kurtosis analysis of the simulated data to monitor the relative performance in recovering the input density parameters. 

The mass slopes corresponding to the input parameters were recovered inside the $95\%$ confidence intervals for the MCMC output in seven out of eight dispersion-only and dispersion-kurtosis analyses for different data sets which demonstrates that the likelihood developed in Section 4 provides an accurate account of the uncertainties. Whilst the outlier for the dispersion-kurtosis is marginally excluded the dispersion-only outlier completely excludes the input core parameters. This anomaly is however rectified with the joint analysis which emphasises that an additional reference to the data can help to reduce spurious sampling effects. As expected the anisotropy parameter $\beta$ was more tightly constrained for all data sets.

In six out of eight data sets the constraints on the mass slope are tighter for the joint analysis with one of these exceptions being the dispersion-only outlier discussed above. For the most realistic cases where the limited sampling caused a scatter that completely obscured the anisotropy and mass slope with a dispersion-only analysis (B, D and  E) the inclusion of the kurtosis was particularly effective and the degeneracy between the extended and vanishing core parameters was broken. Whilst a detailed investigation with other parameterisations is required to confirm this finding it provides strong motivation for further study into the higher order Jeans analysis and demonstrates that even a freely varying fourth order model could prove more constraining than a second order analysis that makes an assumption without reference to the fourth order data.

Ideally we would like to test how constraining the additional fourth order information is in the case where we allow the relationship between $\beta$ and $\beta'$ to vary to a greater or lesser extent. It would also be interesting to try and use the results of N-body simulations to motivate physical choices for the relationship between the two anisotropy parameters. We are working on all these issues and hope to present new results for real dwarf spheroidal data sets in the near future.

\section*{Acknowledgments}
TR thanks the KCL Graduate School for support and Matthew Walker for useful discussions.  MF benefited from very important conversations with Riccardo Catena and gratefully acknowledges support from the STFC.

\bibliographystyle{mn2e2}
\bibliography{tranrep}

\appendix
\section{Order Dependence of Moment Ratios in the Isotropic System}
Let's assume that the center of the galaxy lies in the positive z-direction for a sphere with corresponding polar angles $(0\leq\epsilon_1\leq \pi,\;0\leq\epsilon_2\leq2\pi)$ centered at the stellar position such that the radial velocity is treated as $v_z$ and the angular velocities as $v_x$ and $v_y$. We may thus write,
\begin{equation}
v_{\theta} = v\sin\epsilon_1\cos\epsilon_2
\end{equation}
\begin{equation}
v_{\phi} = v\sin\epsilon_1\sin\epsilon_2
\end{equation}   
\begin{equation}
v_{r} = v\cos\epsilon_1
\end{equation}
where $v$ is the total stellar velocity $v^2=v^{2}_{r}+v^{2}_{\theta}+v^{2}_{\phi}$ and we note that $v_{t} = v\sin\epsilon_1$. The moments may thus be expressed as,
\begin{equation}
\overline{v^{2p}_{r}v^{2q}_{t}} = \overline{v^{2(p+q)}\cos^{2p}\epsilon_1\sin^{2q}\epsilon_1}
\end{equation}
such that to calculate the moment ratios at 2nth order it is sufficient in the isotropic case to consider only the angular contribution $\overline{\Omega}=\overline{\cos^{2p}\epsilon_1\sin^{2q}\epsilon_1}$. Performing the average over all solid angles we integrate $d\Omega=-d(\cos\epsilon_1) d\epsilon_2$ with the uniform probability distribution $P(\cos\epsilon_1,\epsilon_2)=1/4\pi$,
\begin{eqnarray}
 \overline{\Omega} &=& \int^{\pi}_{0}\int^{2\pi}_{0} P(\cos\epsilon_1,\epsilon_2) \cos^{2p}\epsilon_1\sin^{2q}\epsilon_1 \sin\epsilon_1d\epsilon_1d\epsilon_2\nonumber \\
 &=& \frac{1}{4\pi}\int^{2\pi}_{0}\int^{1}_{-1}\cos^{2p}\epsilon_1\sin^{2q}\epsilon_1 d\cos\epsilon_1d\epsilon_2\\
 &=& \frac{1}{2} \int^{1}_{-1}X^{2p}(1-X^2)^{q} dX\nonumber.
\end{eqnarray}
The ratio of tangential moments at fixed order is thus,
\begin{equation}
\frac{m_{a,b}}{m_{c,d}} = \frac{a!}{d!}\frac{\Gamma(a+\frac{1}{2})}{\Gamma(c+\frac{1}{2})}
\end{equation}
where we have used the compact notation \eqref{mnot} and the fact that for moments of equivalent order $a+b=c+d$. The expression is simplified further for \textit{adjacent} $(a=c-1,\;b=d+1)$ moment ratios wherein the shift property of gamma functions may be exploited,
\begin{equation}
\frac{m_{p,q}}{m_{p+1,q-1}} = \frac{2q}{2p+1}.
\end{equation}
Radial moment ratios may then be expressed as the product of adjacent moment ratios,
\begin{eqnarray}\label{prodad}
\frac{m_{n-q,q}}{m_{n,0}} &=& \prod^{q-1}_{i=0} \frac{m_{n-q+i,q-i}}{m_{n-q+i+1,q-i-1}} \\ &=& \prod^{q}_{k=1} 
\frac{k}{(n-q-\frac{1}{2}+k)} = \frac{q!}{(n-q+\frac{1}{2})_{q}}
\end{eqnarray}
and we note crucially that the order dependence of the isotropic system is the same as that of the separable augmented density system. This factor will be used to generalise the factor of $1/2$ present in the second order Binney anisotropy parameter.  

\section{Parent Distributions Parametrised by Variance and Kurtosis}

In the course of the analysis we aim to generate velocity samples from distributions with a range of different variance and kurtosis for the purpose of creating simulated data and bootstrap sampling distributions. For this purpose we use the Pearson family of distributions and a superposition of Gaussians both of which have simple analytic moments that make a simple basis for parameterisation.   

\subsection{Pearson Distributions}
To model leptokurtic distributions with super-Gaussian kurtosis $\kappa > 3$ we employ the Pearson type VII distribution function,
\begin{equation}
P_7(v|\alpha,m)=\frac{\Gamma(m)}{\sqrt{\pi}\Gamma(m-\frac{1}{2})\alpha} \left(1 + \frac{v^2}{\alpha^2}\right)^{-m}
\end{equation}
where $\Gamma(x)$ is the Euler gamma function and the shape parameters $\alpha$ and $m$ are related to the variance $\sigma^2$ and kurtosis $\kappa$ of the distribution via,
\begin{eqnarray}
m(\kappa) &=& \frac{5}{2}+\frac{3}{\kappa-3}\;,\;\;\;\; \; \;\;\;\;\;m>\frac{5}{2} \\
\alpha(\sigma,\kappa) &=& \; \sigma \sqrt{2m-3}\;,\;\;\;\; \; \;\;\;\;\;\alpha>\sqrt{2}\sigma \;.
\end{eqnarray} 
As the type VII distribution is limited to the leptokurtic region $\kappa>3$ we use the type II distribution,
\begin{equation}
P_2(v|s,m)=\frac{\Gamma(\frac{3}{2}-m)}{\sqrt{\pi}\Gamma(1-m)s} \left(1 - \frac{v^2}{s^2}\right)^{-m}
\end{equation}
to model platykurtic distributions where the shape parameter $s$ which defines the domain of the distribution $|v| < s $ is defined by
\begin{equation}
s = \sigma \sqrt{3-2m},\;\;\;\;\;s>\sqrt{3}\sigma
\end{equation}
and in this case $m<0$ yields physical distribution functions for kurtosis values $1.8< \kappa < 3 $. The lower limit, as the kurtosis of the uniform distribution, is a natural lower bound for a measure of peakedness.
In summary, we define the Pearson family as,
\begin{equation}
\mathcal{F}(v|\sigma,\kappa) = \left\{\begin{array}{cc} P_7(v|\alpha,m) & \kappa \geq 3.05 \\ \mathcal{N}(v|\sigma) & 2.95<\kappa<3.05 \\ P_2(v|s,m) & 1.8 < \kappa \leq 2.95 \end{array}\right.
\end{equation}
where $\mathcal{N}$ is the normal distribution. 

 To generate samples from these distributions we first calculate the cumulative distribution function $\Phi$ and then invert it via a look up table for $\Phi^{-1}$. A given velocity that belongs to the distribution is then drawn by generating a random number $x$ from the uniform interval [0,1] and choosing $v= \Phi^{-1}(x)$. Analytic forms for the cumulative distribution functions of the Pearson family are,
\begin{equation}
\Phi_7(v|\alpha,m) = \frac{1}{2} + \frac{v}{\alpha B(m-\frac{1}{2},\frac{1}{2})} . _2F_1\left(\frac{1}{2},m,\frac{3}{2},-\frac{v^2}{\alpha^2}\right)
\end{equation}
 \begin{equation}
\Phi_n(v|\mu=0,\sigma) = \frac{1}{2}\left[1 +\rm{Erf}\left(\frac{v}{\sqrt{2}\sigma}\right)\right]
\end{equation}
\begin{equation}
\Phi_2(v|s,m) = \frac{1}{2} + \frac{\Gamma(\frac{3}{2}-m)v}{\sqrt{\pi}\Gamma(1-m)s}. _2F_1\left(\frac{1}{2},m,\frac{3}{2},\frac{v^2}{s^2}\right)
\end{equation}
where $_2F_1(a,b,c,d)$ is the Gaussian hypergeometric function.

\subsection{Gaussian Superpositions}

To test the impact of varying the parent sampling distribution on our analysis an additional family of probability density functions is required. With strong evidence for multiple stellar populations in dwarf spheroidals we consider distributions that are superpositions of multiple Gaussians. The Jeans analysis requires symmetric distributions and we aim to build distributions similar to those in \cite{Amorisco} with features such as the central dip and turned in tails that are not represented by the Pearson family. An additional benefit is that Gaussian superpositions do not suffer the finite domain of many platykurtic parameterisations and allows a test of this criticism of the Pearson family.    

Symmetric leptokurtic distributions can be formed from the superposition of two Gaussians with zero mean and different widths $\sigma_2 = R \sigma_1 \equiv R \sigma$, 
\begin{equation}
G_L(v|\sigma,b,R) = \frac{N(v|0,\sigma)+b N(v|0,R \sigma)}{\sqrt{2\pi}\sigma(1+b R)}
\end{equation}
where the relative weighting of the Gaussians, denoted by constant $b$, permits an additional freedom. The moments of the model are simple with variance and kurtosis,
\begin{equation}
\sigma^{2}_{L} = \frac{1+b R^3}{1+b R} \sigma^2
\end{equation}
\begin{equation}
\kappa_L = 3 \frac{(1+b R^5)(1+b R)}{(1+b R^3)^2}.
\end{equation}
As the kurtosis is independent of the individual Gaussian variances then given a family of distributions parametrised by the relative weight $b$ one can map the kurtosis to $R$ and then draw $\sigma$ from the desired variance. 
\begin{figure}
	\centering
		\includegraphics[width=9cm]{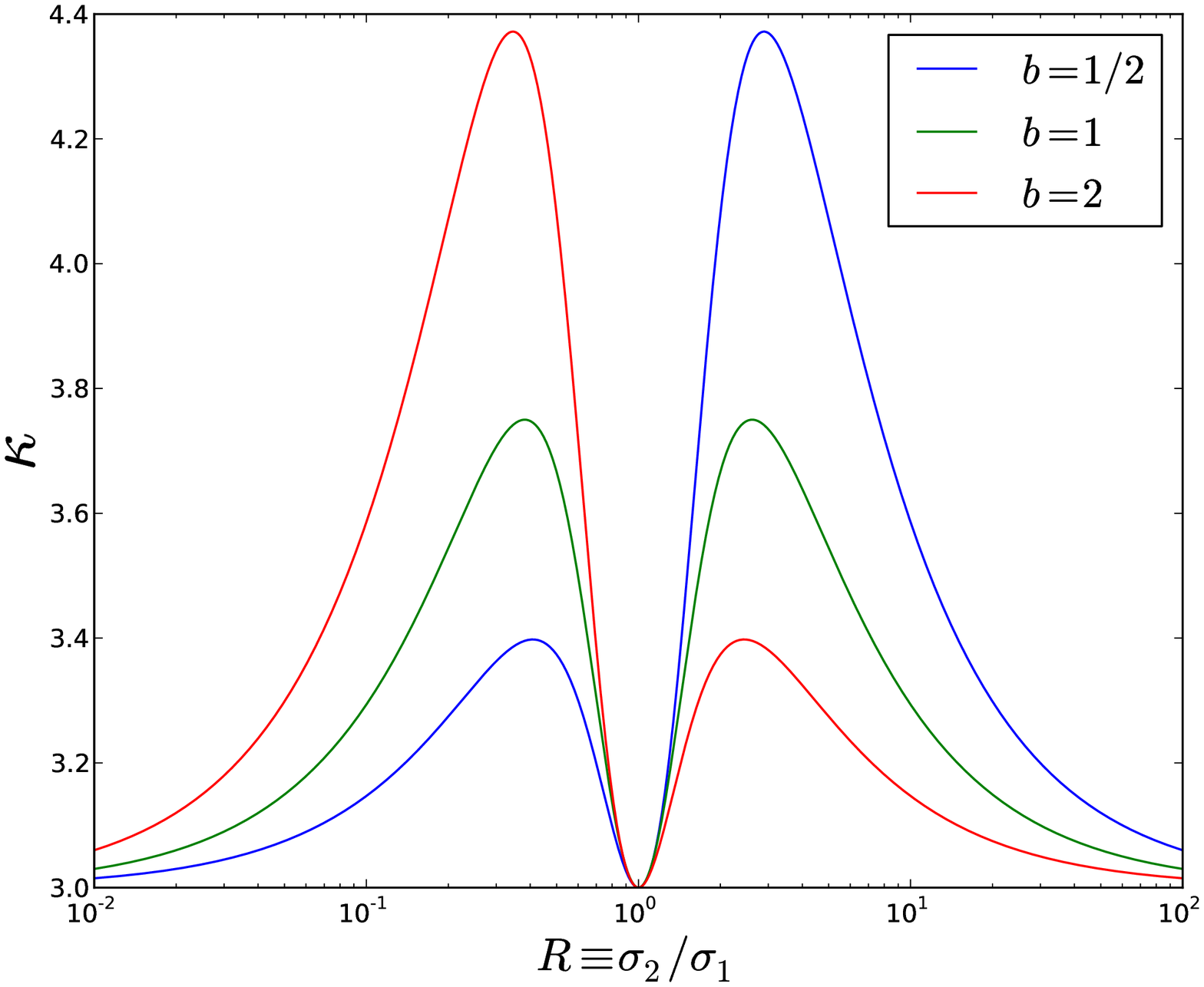}
	
	\caption{Map of kurtosis to R parameter of the leptokurtic family of Gaussian superpositions.} 
	\label{lkR} 
\end{figure}
In Fig. \ref{lkR} we demonstrate the kurtosis map to $R$ for various choices of $b$. At $R=1$ where the variances are equal and as one variance dominates the other the kurtosis collapses to individual Gaussian values of 3 as expected. For all distributions this represents a minimum and thus all distributions of this type are strictly leptokurtic. With $b = 1$ however the maximum at $R=\frac{1}{2}(3 \pm \sqrt{5})$ is limited to $\kappa = 3.75$ so more extreme values of $b$ are required to generate highly leptokurtic samples. One need only consider values $b < 1$ due to the symmetry with the pair $(b,R)$ identical to $(1/b,1/R)$ and we choose the distribution corresponding to the solution closest to $R=1$ which exhibits the least radical departure from the Gaussian line shape. 

For the platykurtic distributions we replace one of the Gaussians with a pair of equally and oppositely displaced ones such that the means are located at $\pm W \sigma$. For simplicity we assume that the variance of the two Gaussians matches that of the Gaussian that remains at the centre. We thus parameterise with the variance scale $\sigma$, the ratio of mean displacement to the variance scale $W$ and a weighting $a$ with,
\begin{equation}
k_P G_P(v|\sigma,a,W) = N(v|0,\sigma)+a\frac{N(v|W\sigma,\sigma)+N(v|-W\sigma,\sigma)}{2}.  
\end{equation}
where $k_p$ is the normalising constant 
\begin{equation}
k_P = \sqrt{2 \pi} \sigma (1+a).
\end{equation}
The variance and kurtosis of this distribution then follow,
\begin{equation}
\sigma^{2}_{P} = (1+\zeta W^2)\sigma^2
\end{equation}
\begin{equation}
\kappa_P = \frac{3+\zeta W^2(6+W^2)}{(1+\zeta W^2)^2}
\end{equation}
where we employ the shorthand $\zeta = a/(1+a)$.

\begin{figure}
	\centering
		\includegraphics[width=9cm]{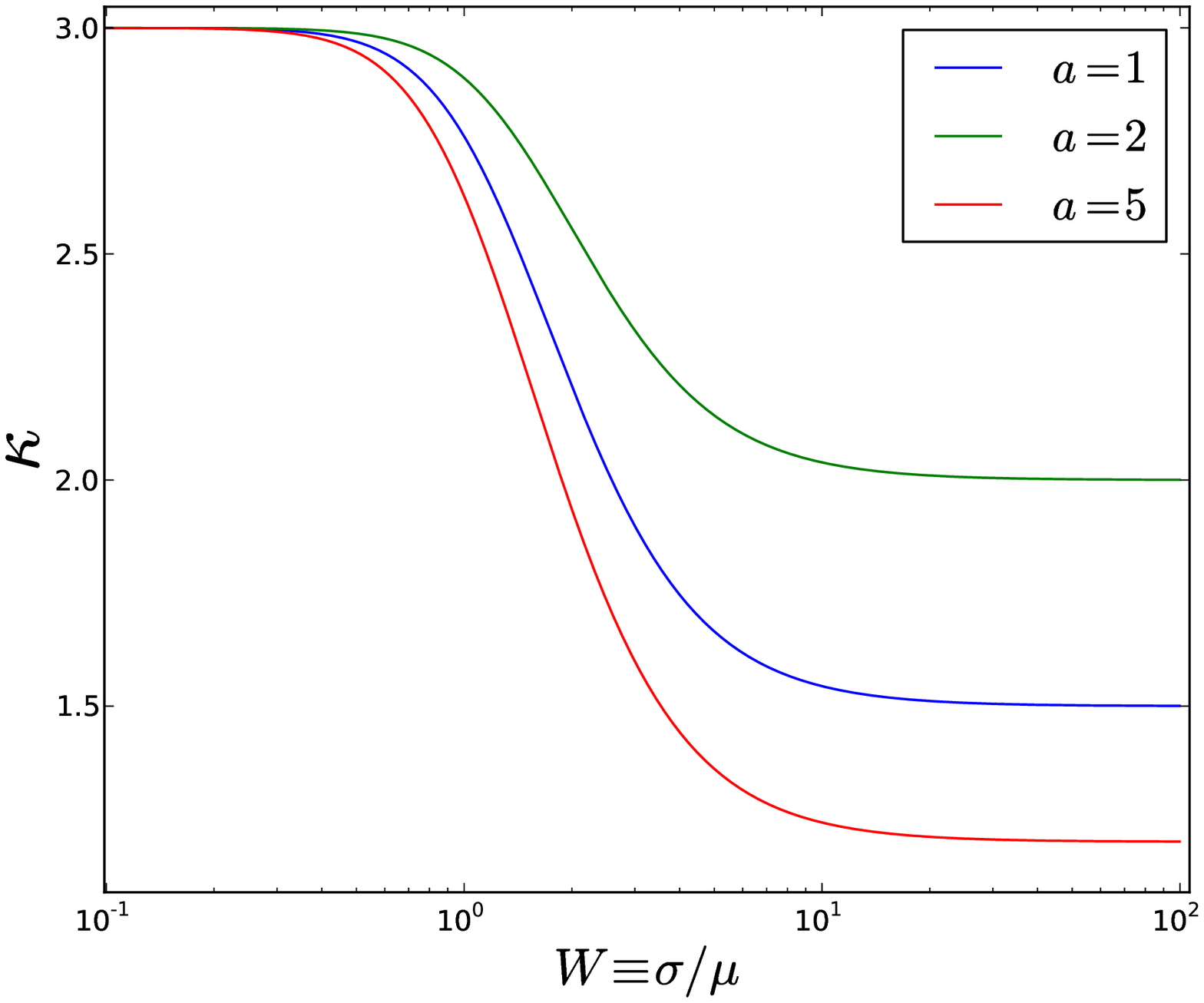}
	
	\caption{Map of kurtosis to W parameter of the platykurtic family of Gaussian superpositions.} 
	\label{pkW} 
\end{figure}
Fig. \ref{pkW}  shows the one to one mapping of kurtosis to W which has the analytic inversion,
\begin{equation}
W^2 = \frac{\zeta(3-\kappa)+\sqrt{\zeta(3\zeta-1)(3-\kappa)}}{\zeta(\zeta\kappa-1)}
\end{equation}
  To close the system we choose $b=0.05$ and $a = (32-9\kappa)/5$ as a small value of $b$ permits large leptokurtosis whilst the more complicated form for $a$ enables both flat topped distributions at low kurtosis and a smooth convergence to the Gaussian distribution as the kurtosis tends to $\kappa=3$. The result which is illustrated in the bottom left of Fig. \ref{sampdists} bears a resemblance to those distributions in \cite{Amorisco} which stem from the constant anisotropy phase space distribution of the isothermal sphere.

To sample from the distributions we derive the cumulative distribution functions,
\begin{equation}
\Phi_L(v|\sigma,b,R) = \frac{1}{2}+\frac{\rm{Erf} \left(\frac{v}{\sqrt{2\pi}\sigma}\right) + b R \rm{Erf}\left(\frac{v}{\sqrt{2\pi}R\sigma}\right)}{2(1+bR)}
\end{equation}
\begin{equation}
\Phi_P(v|\sigma,a,W) = \frac{2\Phi_n(v|0,\sigma)+a\Phi_n(v|W\sigma,\sigma)+a\Phi_n(v|-W\sigma,\sigma) }{2(1+a)}.  
\end{equation}
where $\Phi_n$is the cumulative distribution function of the standard Gaussian.

\end{document}